%% file: paper.tex
\documentclass[fleqn,usenatbib]{mnras}

\usepackage{newtxtext,newtxmath}

\usepackage{amsmath}

\usepackage{pdflscape}
\usepackage{adjustbox}

\usepackage[T1]{fontenc}

\DeclareRobustCommand{\VAN}[3]{#2}
\let\VANthebibliography\thebibliography
\def\thebibliography{\DeclareRobustCommand{\VAN}[3]{##3}\VANthebibliography}

\usepackage{catchfile}

\usepackage{graphicx}	
\usepackage{amsmath}	
\usepackage{fontawesome} 
\usepackage{gensymb}
\usepackage{multirow} 
\usepackage{enumitem}

\defcitealias{2019MNRAS.489.3625C}{Paper I}

\newcommand{\aref}[1]{\hyperref[#1]{Appendix~\ref{#1}}}

\title[Kinematic age of Sco-Cen]{Chronostar. II. Kinematic age and substructure of the Scorpius-Centaurus OB2 association}

\author[M. {\v Z}erjal et al.]{
Maru{\v s}a {\v Zerjal},$^{1,2,3,4}$\thanks{E-mail: mzerjal@iac.es}
Michael J. Ireland,$^{1}$ Timothy D. Crundall,$^{5}$ Mark R. Krumholz,$^{1,6}$ 
\newauthor 
Adam D. Rains$^{1}$
\\
$^{1}$Research School of Astronomy \& Astrophysics, Australian National University, ACT 2611, Australia\\
$^{2}$Centre for Astrophysics, University of Southern Queensland, Toowoomba, Queensland 4350, Australia\\
$^{3}$Instituto de Astrof{\'{\i}}sica de Canarias, E-38205 La Laguna, Tenerife, Spain \\
$^{4}$Universidad de La Laguna, Dpto. Astrof{\'{\i}}sica, E-38206 La Laguna, Tenerife, Spain \\
$^{5}$I. Physikalisches Institut, Universit{\" a}t zu K{\" o}ln, Z{\" u}lpicher Str. 77, D-50937 K{\" o}ln, Germany\\
$^6$ARC Centre of Excellence for Astronomy in Three Dimensions (ASTRO-3D), Canberra ACT 2611, Australia
}

\date{Accepted XXX. Received YYY; in original form ZZZ}

\pubyear{2021}

\begin{document}
\label{firstpage}
\pagerange{\pageref{firstpage}--\pageref{lastpage}}
\maketitle

\begin{abstract}
The nearest region of massive star formation -- the Scorpius-Centaurus OB2 association (Sco-Cen) -- is a local laboratory ideally suited to the study of a wide range of astrophysical phenomena. 
Precision astrometry from the Gaia mission has expanded the census of this region by an order of magnitude. However, Sco-Cen's vastness and complex substructure make kinematic analysis of its traditional three regions, Upper Scorpius, Upper Centaurus-Lupus and Lower Centaurus-Crux, challenging.
Here we use \textsc{Chronostar}, a Bayesian tool for kinematic age determination, to 
carry out a new kinematic decomposition of Sco-Cen using 
full 6-dimensional 
kinematic data. Our model 
identifies
8 kinematically distinct components consisting of 8,185 stars distributed in dense and diffuse groups, each with an independently-fit kinematic age; we verify that these kinematic estimates are consistent with isochronal ages.
Both Upper Centaurus-Lupus and Lower Centaurus-Crux are split into two parts. The kinematic age of the component that
includes
PDS\,70, one of the most well studied systems currently forming planets,
is 15$\pm$3\,Myr. 

\end{abstract}





\section{Introduction}

The closest site of a massive star formation to the Sun is Scorpius-Centaurus OB2 association (Sco-Cen), which spans a large area in the sky due to its proximity. 
The number of confirmed Sco-Cen members has grown over time, from just 512 stars in 1999 \citep{1999AJ....117..354D} to almost 15,000 candidate objects in 2019 \citep{2019A&A...623A.112D} thanks to the precision astrometry from the Gaia mission \citep{2016A&A...595A...1G,2018A&A...616A...1G,2021A&A...649A...1G} that has enabled a more detailed view of the association. For example, \citet{2018A&A...614A..81R} report on a new dense subcomponent V\,1062\,Sco with ages of less than 10 to about 25\,Myr at a distance of 175\,pc.

Historically, Sco-Cen has been divided into three groups, namely Upper Scorpius (USCO), Upper Centaurus-Lupus (UCL) and Lower Centaurus-Crux (LCC), although \citet{2011MNRAS.416.3108R} argued that this split is arbitrary, especially for UCL and LCC. \citet{2018MNRAS.476..381W} suggest that Sco-Cen was likely born highly substructured, with multiple small-scale star formation events.
As summarised by \citet{2008hsf2.book..235P}, there is evidence that star formation in the USCO was triggered by an event occurring in the adjacent but older UCL group. 
These three groups likely represent the remains of low-density, unbound star-forming regions that began to disperse as soon as the star-forming gas was removed; such unbound regions account for $\sim 90\%$ of all star formation in the Milky Way and similar galaxies \citep{Krumholz19a}, making them ideal testing grounds to study a wide variety of phenomena, from early stellar and planetary evolution, to the initial mass function, to the origin of the Galactic field population.
All of these studies, however, rely at least partly on reliable estimates of the age of the dispersing stellar population.
Among the 
techniques available to provide such estimates,
evolutionary tracks 
are the most reliable for
coeval ensembles of stars. 
However,
model uncertainties,
especially on the low-mass end, combined with unresolved binaries and intrinsic luminosity variability in young stars, often limits the precision of 
this method
(e.g. \citealp{2021ApJ...912..137S}). On the other hand, kinematic ages based on the assumption that associations are gravitationally unbound and expanding
provide a model-free alternative, albeit with a requirement for
very precise measurements of parallaxes, proper motions and radial velocities. 
There have been numerous attempts to
estimate ages by tracing the orbits of stars back through time to the point when they were more concentrated than we observe today
(e.g. LACEwING, \citealp{2017AJ....153...95R}), but even with the precision of Gaia astrometry,
such efforts have generally  yielded age estimates with very large uncertainties, and with poor consistency with alternative age-dating methods.
The
primary exception is
\citet{2020A&A...642A.179M}'s study of the Beta Pictoris association -- their kinematic age of 18.5$^{+2.0}_{-2.4}$\,Myr compares well with the alternative techniques, for example
\citet{2014MNRAS.445.2169M}'s 
combined lithium depletion boundary and isochronal age of 23$\pm$3\,Myr. 

Traceback efforts for Sco-Cen have thus far yielded conflicting results.
Recently, \cite{2021MNRAS.tmp.1862S} traced back the celestial coordinates of the Upper Scorpius members on the basis of their present day proper motions and found 8 distinct kinematic subgroups divided into dense and diffuse populations.
An analysis using Gaia\,DR1 data (\citealp{2016A&A...595A...2G}, $\sim$400 stars) by \citet{2018MNRAS.476..381W} found no evidence that the three Sco-Cen subgroups were more compact in the past, but did confirm that the stars are gravitationally unbound and have a nonisotropic velocity dispersion.
On the other hand, \citet{2020ARep...64..326B} 
estimate a linear expansion coefficient of Sco-Cen to be 39\,$\pm$\,2\,$\mathrm{km\,s^{-1}\,kpc^{-1}}$, based on T\,Tauri stars. 


The lack of consistent results from traceback methods led
\citet[hereafter \citetalias{2019MNRAS.489.3625C}]{2019MNRAS.489.3625C} to propose an alternative approach, implemented in the new code \textsc{Chronostar}, based on tracing proposed origin sites forward and comparing against the present-day stellar distribution; the principle advantage of trace-forward as compared to traceback is that it avoids the explosion of uncertainty that inevitably occurs when one attempts to trace backwards the orbits of stars whose position and velocity include observational uncertainties. In \citetalias{2019MNRAS.489.3625C} we
use \textsc{Chronostar} to obtain a
kinematic age of 17.8$\pm$1.2 Myr for the Beta Pictoris association 
in good agreement with the age estimate of \citet{2014MNRAS.445.2169M}. 
A further advantage of
\textsc{Chronostar}'s approach is that it simultaneously and self-consistently determines both kinematic ages and association memberships, rather than relying on predetermined membership lists (as is the case for traceback methods).
This
allows it
to improve the extraction of the members, especially at the low-mass end where the spread in luminosity is large. 
Moreover, the lack of dependence on predefined membership lists allows \textsc{Chronostar} to discover new kinematic groups blindly, an ability that is particularly important for diffuse groupings whose overdensity might not be apparent in 2D or even 3D position space, but that become apparent in 6D phase space.



In this work we upgrade \textsc{Chronostar} to 
enable efficient processing of larger data sets than in \citetalias{2019MNRAS.489.3625C}
(\autoref{sec.chronostar_improvements}) and use it to explore the kinematic substructure of the Scorpius-Centaurus association (\autoref{sec.fitting_procedure}). 
\autoref{sec.discussion} addresses the individual components while \autoref{sec.age_validation} validates their kinematic age. We present our conclusions in \autoref{sec.conclusions}.


\section{\textsc{Chronostar}} \label{sec.chronostar_improvements}

We first summarise the \textsc{Chronostar}\footnote{\url{https://github.com/mikeireland/chronostar}} algorithm, referring readers to \citetalias{2019MNRAS.489.3625C} for a full description and a list of tests we have performed to validate the code. We then describe improvements made to the code since publication of \citetalias{2019MNRAS.489.3625C}, which we use here.

\subsection{Summary of the \textsc{Chronostar} algorithm}

\textsc{Chronostar} is a robust Bayesian method for kinematic age determination of young stellar associations. Its autonomous examination of the parameter space 
addresses the complex substructure of associations, their membership determination and evolution of the associated models. We model an association as consisting of one or more independent components, each of which is described by an age and an initial (i.e., at age zero) distribution in the 6D Cartesian phase space XYZUVW. The spatial part of the coordinate system is centered at the Sun’s projection on the Galactic plane, while the origin of the velocity is placed in the Local Standard of Rest \citep{2010MNRAS.403.1829S}.
For simplicity, the initial distribution of each component is approximated as an isotropic Gaussian, with the same standard deviation in all three spatial dimensions and all three velocity dimensions. In our model, time 0 corresponds to the time at which the stars transition from being gravitationally bound to moving on ballistic orbits through the Galaxy. In order to obtain the \textit{present-day} distribution of stars in XYZUVW space, we trace each component \textit{forward} in time to the present day. We compute the orbits using \textsc{galpy} \citep{2015ApJS..216...29B}, with \texttt{MWPotential2014} for the Galactic potential. 

In order to fit the parameters describing each component (its kinematic age, origin point in 6D phase space, and initial dispersion in position and velocity), we compare the present-day distribution that results from the trace-forward to the observed distribution of members of that component in 6D phase space. In turn, we identify those observed stars that are most likely to be members of each component based on the overlap integral between the Gaussian describing the present-day distribution of that component and the Gaussian describing the error distribution of each prospective member star in 6D phase space, including the full covariances of the observational errors. The circular dependency -- the need to know the members in order to fit the parameters describing each component, and the need to know the parameters of each component in order to determine which stars are members -- is resolved using the Expectation-Maximisation (EM) algorithm. This iterative algorithm starts from an initial guess of membership assignments, then fits the properties of each component for those assignments. Once the best-fitting component parameters have been found, the algorithm then fixes those parameters and recomputes the membership probabilities for each observed star. The algorithm then iterates until the parameters of each component and the membership assignments to them converge. The important advantage of this approach is that, because the membership lists are not fixed \textit{a prori}, \textsc{Chronostar} can blindly discover new moving groups, study the substructure of known associations, and revise its membership census. This in turn enables the accurate determination of kinematic ages.

In addition to the EM step, \textsc{Chronostar} adds a third layer of iteration. The EM algorithm operates on a fixed number of components, but we do not know \textit{a priori} how many components we should use to fit a given data set.
To handle this problem, \textsc{Chronostar} starts assuming a single component, and after the EM algorithm converges, we split the resulting component into two components, then rerun the EM algorithm for those two components. We then compute the Bayesian information criterion (BIC), which represents the likelihood function penalised by the number of parameters in the model, for the one-component and two-component fits; if the BIC for one-component fit is preferred we stop, while if the two-component fit is preferred, we repeat the procedure, splitting the two components into three, and so forth, stopping once the BIC indicates that adding a further component worsens rather than improves the model.

\subsection{Improvements}

The version of \textsc{Chronostar} described in \citetalias{2019MNRAS.489.3625C} has a number of computational bottlenecks that make it impractical to apply to the larger data sets required to model a large association such as Sco-Cen. First, the process of splitting groups into subgroups means that the total number of single component fits is 1+$\sum_{i=0}^{N_c}{(N_c-i)(N_c-i-1)}$, 
where $N_c$ is the total number of final components; for large $N_c$ this sum scales as $N_c^3$. Second, the time required for each component fit is proportional to the number of stars. 

To overcome these limitations, we have made the following key improvements to \textsc{Chronostar}, which do not affect the underlying formalism, but significantly improve speed:

\begin{itemize}
    \item When carrying the likelihood maximisation step in order to find the best-fitting parameters for each component, we have replaced the Markov chain Monte Carlo method used in \citetalias{2019MNRAS.489.3625C} with a simple derivative-free gradient descent algorithm \citep{NelderMead}; this locates the maximum significantly more rapidly.
    \item When fitting multiple components, we parallelise the fit using multiple CPUs, so that we fit each component simultaneously using its own thread.
    \item The version of the code described in \citetalias{2019MNRAS.489.3625C} integrated orbits (required for the trace-forward step) numerically. For ages $<30$ Myr, we replace this numerical integration with an analytic epicyclic approximation. We describe this in more detail in \aref{sec.epicyclic}.
\end{itemize}

\begin{figure}
\includegraphics[width=\linewidth]{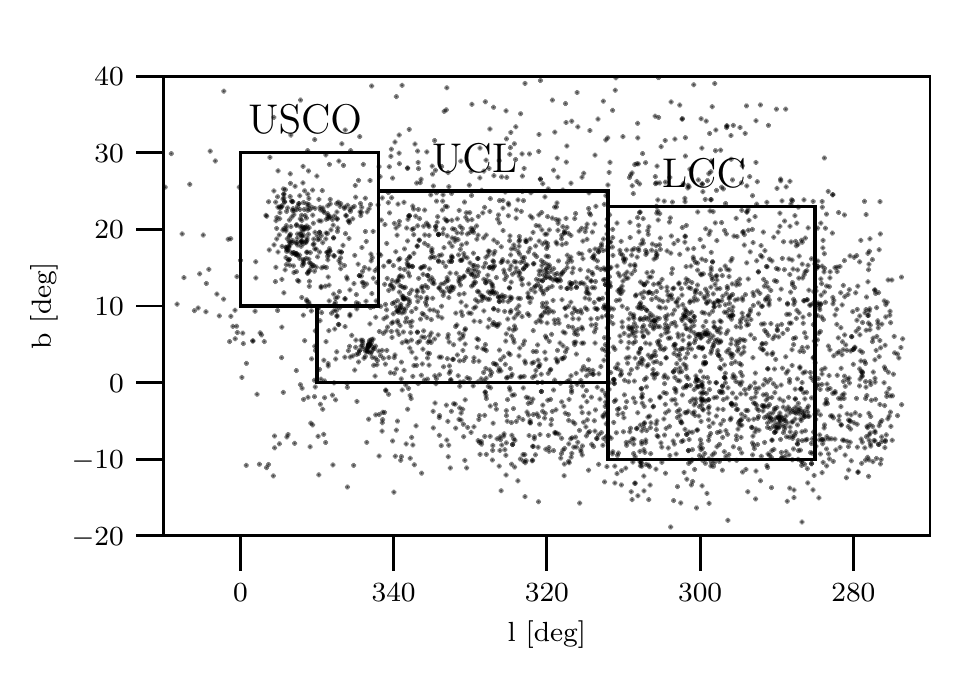} 
\caption{
Distribution in $(l,b)$ of the 3,591 stars used in our initial fit of three subgroups (\autoref{ssec.subgroup}). The three boxes indicate the traditional boundaries of the Upper~Scorpius (USCO), Upper~Centaurus-Lupus (UCL) and Lower~Centaurus-Crux (LCC) subgroups.
\href{https://github.com/mikeireland/chronostar/blob/master/projects/scocen/galaxy_input_data_with_RVs_small.py}{\faGithub}
}
\label{fig.gx_input}
\end{figure}

\section{Fitting Sco-Cen Components} \label{sec.fitting_procedure}



The process of fitting the $N_c$ components and assigning membership probabilities to all stars in the vicinity of Sco-Cen has several steps. Aspects of this procedure are expected to change for future papers with even more speed improvements (e.g. further moving slow aspects of python code to C), but the general procedure for applying  \textsc{Chronostar} to a new association or region of the sky remains as follows:

\begin{enumerate}
    \item Based on prior knowledge, define a region of 6-dimensional kinematic space and specify data quality requirements in order to define a sample of stars.
    \item After an initial fit, refine the association boundaries and initial conditions, and potentially refine the input catalog. 
    \item Without re-fitting component properties, expand the input list of stars to those with lower quality data, and beyond the original 6-dimensional region.
\end{enumerate}
The reasons for step (iii) are, first, that components inevitably extend beyond the original 6-dimensional region, so in order to include all their members we must extend them. Second, we have found that fitting components to stars with poor quality data increases both computational expense and the risk of producing artefacts; it is substantially more efficient to carry out an initial fit to determine component properties using only high quality data, and then extend the membership list to lower-quality data without re-fitting the individual components.

For the fitting in this paper, we now describe the process as carried out for Sco-Cen.

\subsection{Subsample Definition and Subgroup Fits}
\label{ssec.subgroup}


Our initial sample was based on the Sco-Cen members from the BANYAN list \citep{2018ApJ...856...23G}, which was based on the traditionaly boundaries of USCO, UCL and LCC. We drew a 6D Cartesian box around all the stars in three subgroups, adding a margin of 30\,pc and 5\,km\,s$^{-1}$ on each side and took all stars within those boxes from the 6D Gaia catalog to make sure no new potential members were rejected. Only 3,591 stars with 6D kinematic data from Gaia were included, with a 2-dimensional distribution in $l$ and $b$ shown in \autoref{fig.gx_input}. 

The three subsets were fitted separately using \textsc{chronostar}. The kinematic fits resulted in 22 unique components altogether, many of which account for non-homogeneous background. Some components were discovered twice in neighboring associations since the input datasets were slightly overlapping due to the wide cut-off boundaries. 

\subsection{Refined Full-Association Fitting}

To refine the fit, we merged together these 3,591 stars and unique components from the three initial subsets and run a second \textsc{chronostar} iteration on this limited subsample of stars. This fit was continued until completion, i.e., until the Bayesian Information Criterion could not be improved by splitting any of the components further. 



The refined fit converged with 19 components, where 6 of them again serve as background, and 6 of them are complex, meaning that they contain both old and young stars. The remaining 7 components all have birth spatial dispersion less than 8\,pc and velocity dispersion less than 1.1\,km\,s$^{-1}$. The average velocity dispersion of 0.7\,km\,s$^{-1}$ is consistent with velocity dispersions inferred from spatial distributions of young stars \citep{2008ApJ...686L.111K}. These 7 components are further subdivided into diffuse and dense components based on their birth velocity dispersion, which also correlates with their angular distribution on the sky. We present the fitted parameters in \autoref{tab.components_fit}. Given the large number of parameters and that the split into sub-associations is not likely to be unique, we do not derive formal uncertainties of the birth 6-dimensional positions or kinematic widths $\sigma$. 

It is clear that in the future, expanding the input list of stars beyond the 3,591 initially selected with full 6D Gaia kinematics may result in different and more complex fits. We make no orbital corrections for binary stars in this list. We further emphasize that our data selection is based purely on position in 6D parameter space, and completely neglects stellar age, location in the color-magnitude diagram, or any other method of selecting young stars. For these reasons, we anticipate that future work will find significantly more substructure. However, these 13 non-background components still represent a rich data set to explore, especially when additional potential members are added from stars beyond the initial sample of 3,591.

\newcommand{\componentsfit}{\input{componentsfit.tex}}
\newcommand{\componentsfitbackground}{\input{componentsfitbackground.tex}}
\newcommand{\componentsfitcomplex}{\input{componentsfitcomplex.tex}}
\newcommand{\componentsfitgood}{\input{componentsfitgood.tex}}
\newcommand{\componentsfitdense}{\input{componentsfitdense.tex}}
\begin{landscape}
\begin{table*}
\hspace*{-8cm}
\begin{tabular}{l | rrrrrrrrrrrr | rrrrrrrrrrr | l }
ID & $X_\mathrm{t}$ & $Y_\mathrm{t}$ & $Z_\mathrm{t}$ & $U_\mathrm{t}$ & $V_\mathrm{t}$ & $W_\mathrm{t}$ & $\sigma_{X_\mathrm{t}}$ & $\sigma_{Y_\mathrm{t}}$ & $\sigma_{Z_\mathrm{t}}$ & $\sigma_{U_\mathrm{t}}$ & $\sigma_{V_\mathrm{t}}$ & $\sigma_{W_\mathrm{t}}$ & $X_0$ & $Y_0$ & $Z_0$ & $U_0$ & $V_0$ & $W_0$ & $\sigma_{X_0}$ & $\sigma_{V_0}$ & $T_\text{age}$ & $\Delta T_\text{birth}$ & N$_\mathrm{fit}$  \\
 & pc & pc & pc & $\mathrm{km\,s^{-1}}$ & $\mathrm{km\,s^{-1}}$ & $\mathrm{km\,s^{-1}}$ & pc & pc & pc & $\mathrm{km\,s^{-1}}$ & $\mathrm{km\,s^{-1}}$ & $\mathrm{km\,s^{-1}}$ & pc & pc & pc & $\mathrm{km\,s^{-1}}$ & $\mathrm{km\,s^{-1}}$ & $\mathrm{km\,s^{-1}}$ & pc & $\mathrm{km\,s^{-1}}$ & Myr & Myr &  \\
 \hline
  \multicolumn{3}{c}{Diffuse components} & \\
 \input{componentsfitgood.tex}
 \hline
 \multicolumn{3}{c}{Dense components} & \\
 \input{componentsfitdense.tex}
 \hline
 \multicolumn{3}{c}{Complex components} & \\
 \input{componentsfitcomplex.tex}
 \hline
 \multicolumn{4}{c}{Background components} & \\
 \input{componentsfitbackground.tex}
\end{tabular}
\caption{Components in the Sco-Cen model today (time t) and at birth (time 0). All velocities are with respect to the local standard of rest (see \autoref{sec.chronostar_improvements}) and not the Sun. Components are split into four different types based on their structure and stellar population. Velocity dispersion is, with a few exceptions, a good indicator of the nature of the component. Diffuse components on average have larger velocity dispersions than dense components; dense components are likely gravitationally bound.
Complex components encompass multiple populations (including pre-main sequence stars) and/or a significant substructure, and need a further split. Columns $T_\text{age}$ and $\Delta T_\text{birth}$ provide kinematic age and crossing-time while N$_\mathrm{fit}$ gives a number of stars used in the fit. 
Full covariance matrices are available online. 
\href{https://github.com/mikeireland/chronostar/blob/master/projects/scocen/print_components_table_for_paper.py}{\faGithub}
\href{https://github.com/mikeireland/chronostar/blob/master/projects/scocen/data/covariance_matrices_for_comps_for_upload_to_the_journal.py}{\faGithub}
}
\label{tab.components_fit}
\end{table*}
\end{landscape}


\subsection{Expanded candidate list and membership probabilities} \label{sec.candidates}
For the third step in our fitting procedure, we prepared a significantly new catalog of candidate stars with lower-quality data and extending significantly beyond the range of our initial fit, but did not re-fit the components with this larger input data set. 
The candidate stars were selected from the Gaia\,DR2 catalog with the following requirements:
5\,mas\,$<$\,\texttt{parallax}\,$<$\,12\,mas (corresponding to distances between 80 and 200\,pc), \texttt{parallax\_error}\,$\le$\,0.3\,mas which assures parallax errors smaller than 6\% at 200\,pc, \texttt{l}\,$>$\,240{\degree} or \texttt{l}$\,<$\,40{\degree}, and {$-60$\degree\,$<$\,\texttt{b}\,$<$\,70\degree}. 
Such selection results in 641,889  
candidate stars, 127,228 
of which have full 6D measurements available in the Gaia\,DR2 catalog. This relatively small number of stars with full 6D measurements is because Gaia had a relatively bright magnitude limit for the radial velocity spectrograph, and also was not able to provide velocities for particularly hot or cool stars. 

In order to maximise the probability of selection of potential members, this list was cross-matched with the external spectroscopic surveys via Gaia\,DR2 designation \texttt{source\_id} to obtain complementary radial velocity measurements. 
In total, we took measurements for 11,066 stars from the GALAH\,DR3 survey \citep{2021MNRAS.506..150B}, 6,554 from the RAVE\,DR6 list \citep{2020AJ....160...82S}, 4,214 stars from the APOGEE\,DR16 list \citep{2020ApJS..249....3A}, 191 stars as tabulated in the Banyan\,$\Sigma$ paper \citep{2018ApJ...856...23G},
and 168 stars from \citet{2021MNRAS.503..938Z}. 
If the star had multiple measurements in one of the external catalogs, we took the median value.
In case the star had been found in more than one catalog, the value with the smaller uncertainty was selected. Therefore, 15,597 radial velocities from the Gaia catalog were replaced, and in total, 22,193 sources from the external catalogs are used. In total, 133,824 stars from our candidate list have a full 6D information available (21\%).
As our fast overlap integral calculation requires a non-singular covariance matrix, we set the missing radial velocity values to 0$\pm$10,000\,$\mathrm{km\,s^{-1}}$. Such large uncertainties result in the stellar covariance matrices resembling a highly elongated ellipsoids and give membership probabilities indistinguishable from those with formally missing radial velocities.

In the final step we compute the overlap integrals between the components and the stars themselves. Essentially, this is the expectation step from \textsc{Chronostar}'s algorithm used with the new larger dataset. These overlaps are then normalised such that the sum of the membership probabilities for each star equals unity. This takes into account all components from the model, including background components.
A full list of stars with their membership probabilities greater than 50\% to all non-background components is provided in \autoref{tab.results} together with their radial velocities.

The distribution of membership probabilities for each of the non-background components typically has a U-shape, with most of the stars having probabilities either close to 0 or close to 1. 

\subsection{Stellar Age uncertainties}

Age uncertainties $\Delta T_\text{birth}$ as listed in  \autoref{tab.components_overlaps} are determined as the ratio of $\sigma_{X_0}$ and $\sigma_{V_0}$ from \autoref{tab.components_fit}, which represents a kinematic crossing time  for stars within the association at birth. Individual stars in the association are not expected to have ages that are more uniform than $\Delta T_\text{birth}$, unless the stars are part of sub-associations.

Seven of the components have birth crossing times shorter than their age, which suggests that the age estimates are reliable. For these components, the velocity dispersions at birth are also similar to isothermal sound speeds in young associations, meaning that these age uncertainties provide a realistic estimate for the dynamical timescales over the spatial scales defined by the associations. More refined kinematic estimates could only be obtained by further resolving substructure.

Several dense components (D, H and I) are potentially gravitationally bound. This contradicts \textsc{Chronostar}'s assumption that stars move ballistically through the Galaxy, and means that our kinematic ages are likely to be underestimated. Physically, this is simply because a bound group of stars remain coherent in phase space far longer than would be expected for stars on ballistic orbits; since we do not account for this effect, \textsc{Chronostar} interprets tight clustering in phase space as young age. Consistent with this analysis, the kinematic ages we obtain for components D, H, and I are much smaller than the values from the alternative techniques such as isochronal fitting and lithium depletion boundary age and have large uncertainties. They are not reliable estimates, but are nonetheless reported for completeness.

\section{Results} \label{sec.discussion}

\begin{figure*}
\includegraphics[width=\linewidth]{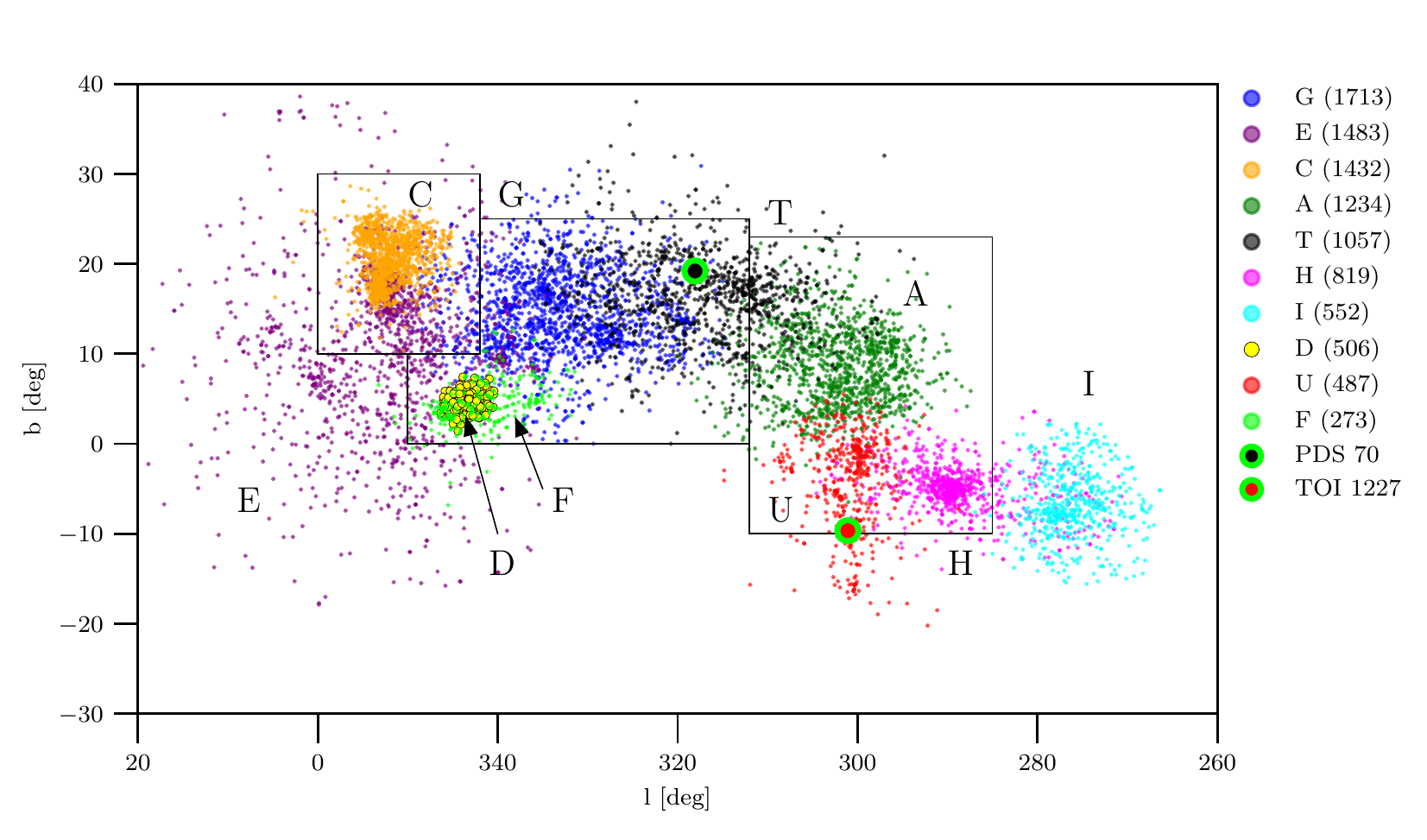} 
\caption{Components of the Scorpius-Centaurus association and its immediate neighbourhood, as described with the kinematic model from \textsc{Chronostar}. There are 9,556 stars in this plot, which shows all stars with membership probabilities at least 50\% in one of the 10 components indicated. The numbers of members for each component are shown in the legend; stars are assigned to the component for which the membership probability is highest. Black lines denote the traditional split into the three regions of the association USCO, UCL and LCC. Exoplanetary hosts PDS\,70 and TOI\,1227 are members of components T and U, respectively.
\href{https://github.com/mikeireland/chronostar/blob/master/projects/scocen/galaxy_components_paper.py}{\faGithub}
}
\label{fig.gx}
\end{figure*}

Our results from \textsc{Chronostar} provide a kinematic decomposition of Sco-Cen that does not rely on any kinematic or evolutionary filters.
The kinematic model that best describes the input 6D dataset of the Scorpius-Centaurus region in this work is composed of 13 non-background components. 
We plot the $(l,b)$ positions of all identified stars with membership probabilities of more than 50\% in \autoref{fig.gx}; colour indicates the highest-probability component assignment for each star. The Figure reveals the familiar outline of the association including its complex substructure with both dense and diffuse groups, but also includes components that go beyond the traditional boundaries of USCO, UCL and LCC (e.g. component E) and that include clusters (IC\,2602 - component H and Platais\,8 - component I). Component populations from a few 100 (component F) to almost 2,000 (component G), and many of them feature further substructure. The total number of stars in each component with membership probability greater than 0.5 is presented in  \autoref{tab.components_overlaps}. 


\newcommand{\componentsoverlaps}{\input{componentsoverlaps.tex}}
\begin{table}
\begin{center}
\begin{tabular}{r|@{\qquad}rrr|@{\qquad}rr}

ID & \multicolumn{3}{c}{Members} & $ T_\text{age}$ & $\Delta T_\text{birth}$ \\
 & p$>$0.5 & p$>$0.8 & p$>$0.9 & Myr & Myr \\
 \hline
\multicolumn{3}{l}{USCO Components} & & \\
C & 1432 & 1246 & 1106 & 4 & 4  \\
E$^a$ & 1483 & 949 & 678 & 11 & 4  \\
 \hline
\multicolumn{3}{l}{UCL Components} & & \\
D & 506 & 416 & 349 & 2 & 7  \\
F & 273 & 197 & 169 & 15 & 2  \\
G & 1713 & 1023 & 633 & 13 & 8  \\
T & 1057 & 747 & 590 & 15 & 3  \\
B$^a$ & 4385 & 2344 & 1444 & 5 & 11  \\
 \hline
\multicolumn{3}{l}{LCC Components} & & \\
A & 1234 & 846 & 610 & 7 & 5  \\
U & 487 & 408 & 376 & 9 & 4  \\
 \hline
Total & 8185 & 5832 & 4511 & & \\
 \hline
 \hline
\multicolumn{3}{l}{Known clusters} & & \\
H & 819 & 785 & 759 & 15 & 21  \\
I & 552 & 491 & 445 & 2 & 11  \\
\end{tabular}
\end{center}
$^a$Complex, multi-population component.
\caption{Number of members in individual components with membership probabilities above 0.5, 0.8 and 0.9. Total number includes all stars in Sco-Cen components except B, which is composed of young and dominantly old populations of stars.
Components H (IC\,2602) and I (Platais\,8) are not part of the association. 
\href{https://github.com/mikeireland/chronostar/blob/master/projects/scocen/print_components_overlaps_table_for_paper.py}{\faGithub}
}
\label{tab.components_overlaps}
\end{table}


This is the first time kinematic analysis has yielded a reliable extraction of young stars from an unbiased dataset, based solely on their common motion through space.
We emphasize that the majority of stars (80\%) lack radial velocities, but they are nevertheless constrained well enough in the remaining 5 dimensions to yield significant overlaps with the Sco-Cen model components. 
Because there is no kinematic discrimination in the input catalog, \textsc{Chronostar} reconstructs all major kinematic groups in the data regardless of their origin or connection to Sco-Cen. We thus note that not all substructures are necessarily related to the Scorpius-Centaurus association. We discuss the individual components below and validate their kinematic ages in  \autoref{sec.age_validation}. At various points we provide colour-magnitude diagrams for different components; all of these are dereddened, using the reddening correction procedure described in \aref{sec.reddening}.






\subsection{Components G and T: Upper Centaurus-Lupus}
\begin{figure*}
\includegraphics[width=\linewidth]{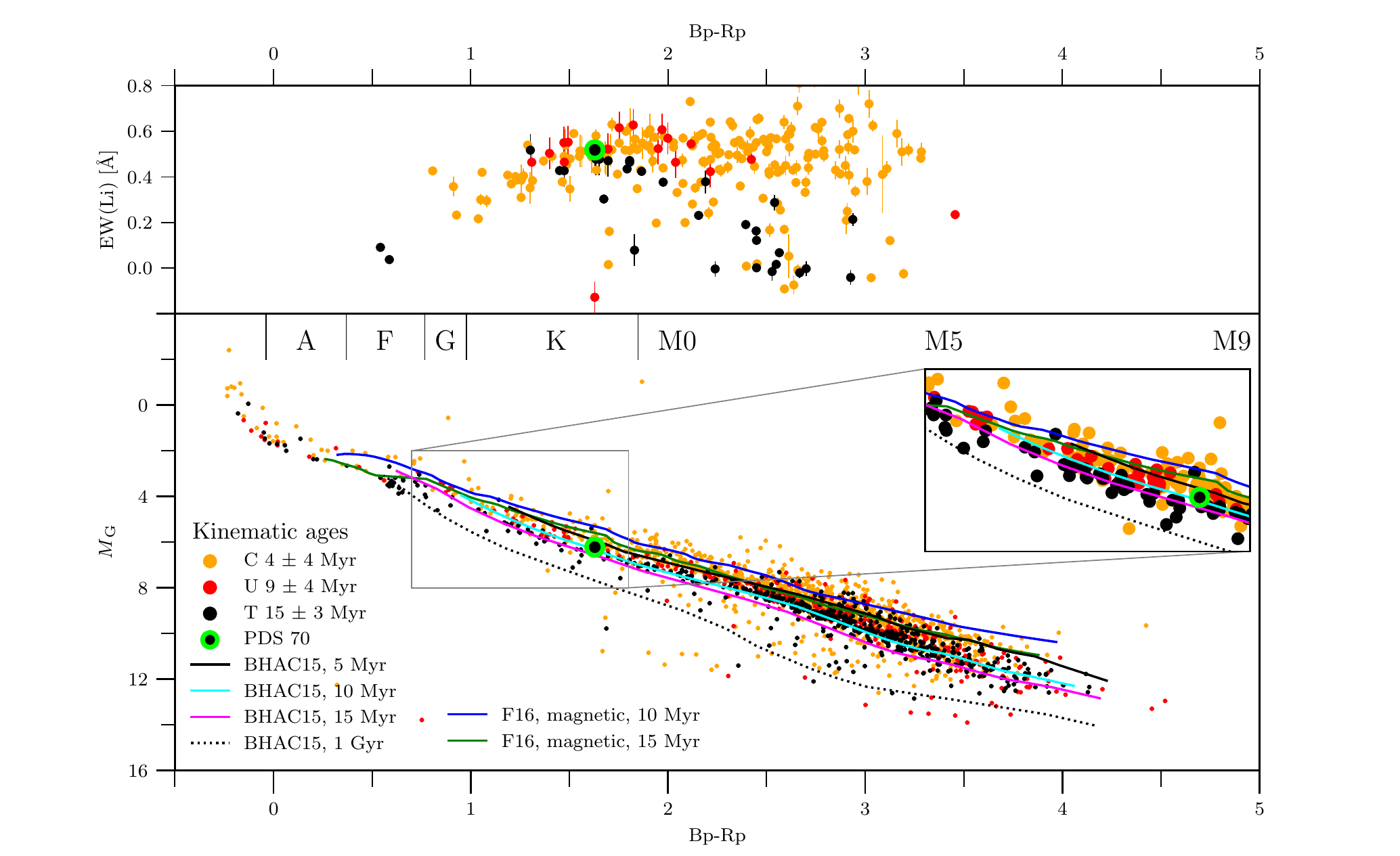}
\caption{\textit{Lower}: A Gaia $M_\text{G} - (B_\text{p}-R_\text{p})$ colour-magnitude diagram of our candidates in components C, U and T with membership probability greater than 90\%.
Component T is kinematically the oldest component ($15\pm3$\,Myr) and shows lower overluminosity and lithium strength than kinematically younger components U and C. The kinematic age of component U is 9$\pm$4\,Myr, but its lithium strength does not show a significant difference with component C; the latter is likely composed of multiple populations with an age spread.
An equal mass binary sequence is clearly present in the component T; the exoplanetary host PDS\,70 belongs to this component. 
For reference, we overplot \citealp{2015A&A...577A..42B} (BHAC15) models and isochrones with magnetic fields from \citealp{2016A&A...593A..99F} 
(F16). 
\textit{Upper}: EW(Li) measurements for candidates as a function of Gaia $B_\text{p}-R_\text{p}$. \href{https://github.com/mikeireland/chronostar/blob/master/projects/scocen/cmd_age_sequence_CUT_with_lithium.py}{\faGithub}
}
\label{fig.cmd_li_CUT}
\end{figure*}

\begin{figure*}
\includegraphics[width=\linewidth]{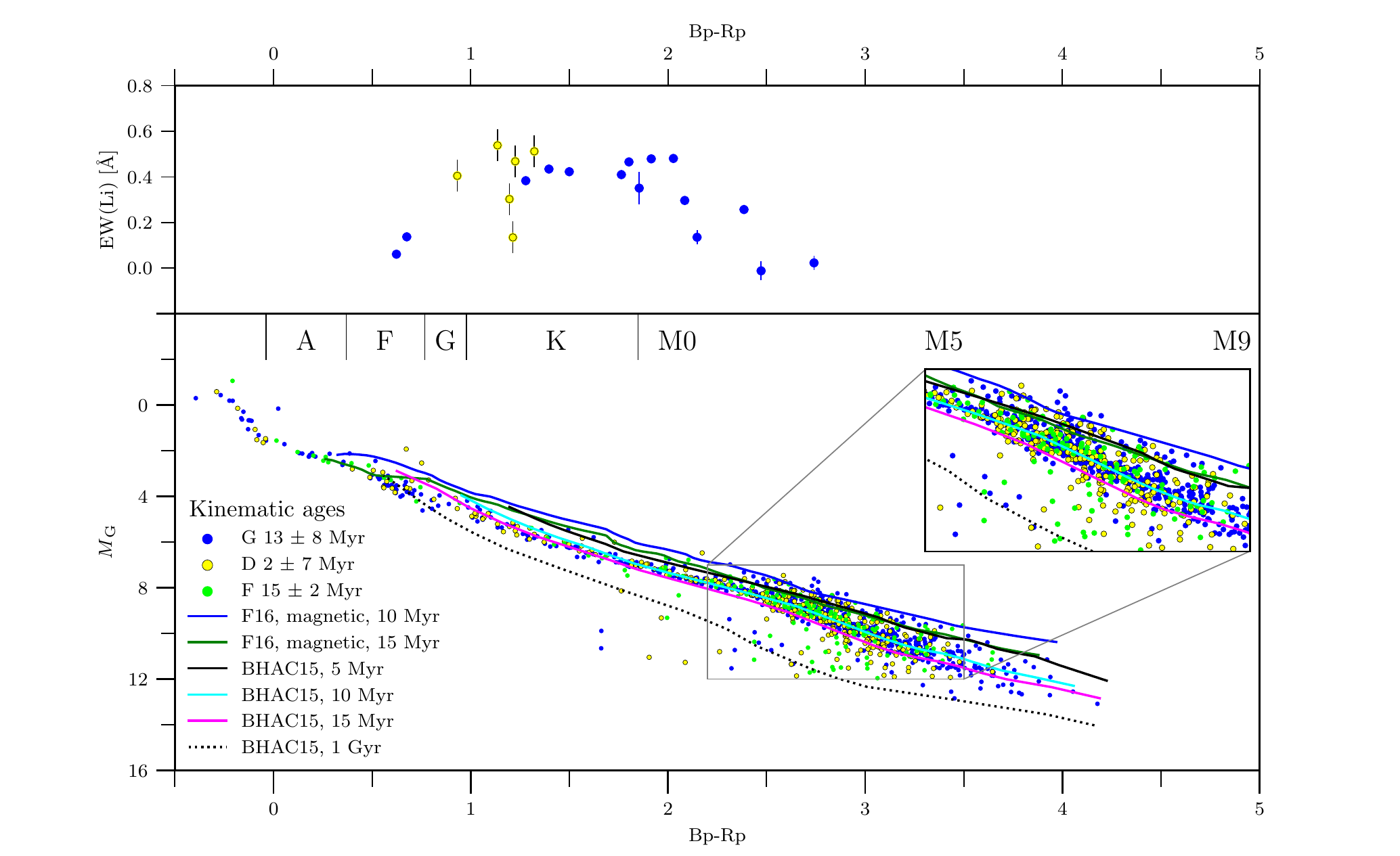}
\caption{\textit{Lower}: A Gaia $M_\text{G} - (B_\text{p}-R_\text{p})$ colour-magnitude diagram of our candidates in components G, D and F with membership probability greater than 90\%. All three components show similar overluminosities.
An equal mass binary sequence is visible in all of them.
\textit{Upper}: EW(Li) measurements for candidates as a function of Gaia $B_\text{p}-R_\text{p}$. 
\href{https://github.com/mikeireland/chronostar/blob/master/projects/scocen/cmd_age_sequence_GDF_with_lithium.py}{\faGithub}
}
\label{fig.cmd_li_GDF}
\end{figure*}

Within the region that corresponds to Upper Centaurus-Lupus, \textsc{Chronostar} identifies two major components, G and T, with 1,713 and 1,057 probable members, respectively.
Their colour-magnitude diagrams (\autoref{fig.cmd_li_CUT} and \autoref{fig.cmd_li_GDF}) show that the stars we identify as belonging to these components lie well above the main sequence, confirming their youth. The Figures also reveal a significant number of equal-mass M dwarf binaries in both components.
Although their spatial distribution shows complexity beyond a simple split into two groups (i.e. the Lupus complex, e.g. \citet{1999PASJ...51..895H} and the two filaments of the T component that continue into the component A), stars of both parts show similar distance distributions (\autoref{fig.distance_distribution_components}), ranging from 100 to 170\,pc with centres at $\sim$135\,pc (T) and $\sim$145\,pc (G).

The shape of both model components taken together is approximately an oblate spheroid with a 15\,pc semi-axis in the Y direction, and 5\,pc in the orthogonal directions.  
While the elongated shape of the UCL alone might have forced \textsc{Chronostar} to split it into two parts, components T and G also show slightly different kinematic properties.
The G component appears to be moving faster towards the Galactic centre since its U velocity at birth (2.5\,$\mathrm{km\,s^{-1}}$) is significantly higher than for the T component (0.2\,$\mathrm{km\,s^{-1}}$). This difference is significant, since the dispersions within each component are only 1 km\,s$^{-1}$. Taken together, these two components could be modelled as a rotating elongated ellipse at birth. 




\subsection{Components A and U: Lower Centaurus-Crux}

Lower Centaurus-Crux is comprised of two kinematically distinct groups A and U, with 1,234 and 487 members, respectively. Component U extends to the south towards the region in the sky where $\epsilon$\,Cha is found at $(l, b)\approx(300,-16)$ deg. 
It includes 3 stars from the \citet{2013MNRAS.435.1325M} list of $\epsilon$\,Cha members.

The model velocity dispersion of the U component (1.0\,$\mathrm{km\,s^{-1}}$) is similar to those of the dense components (\autoref{tab.components_fit}). This is perhaps not surprising as there is an overdensity of stars in the northern part at (l, b)$\approx$(300,\,0)\,deg. \citet{2018ApJ...868...32G} report a new large moving group in the LCC composed of four kinematic subgroups. Their groups A0 and A are fully contained in our component U (195 stars), while only 86\% (698 stars) of the members of their groups B and C were found to belong to \textsc{Chronostar}'s component A. Members of their group Z are distributed between our components A, U and J.

\begin{figure*}
\includegraphics[width=\linewidth]{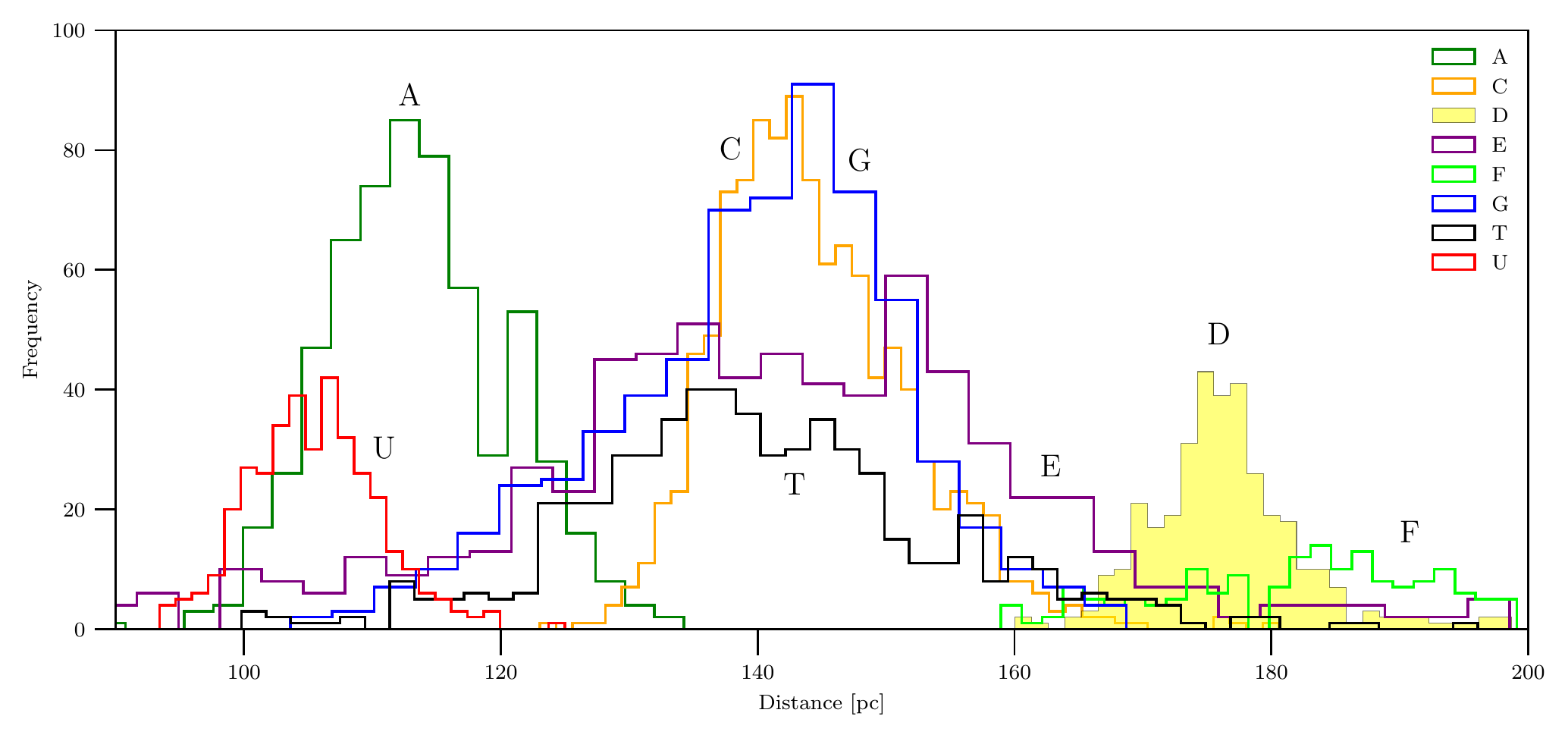}
\caption{Distance distribution for stars with Sco-Cen membership probability p$>$0.9. While components G and T (both in the UCL region) have similar distance distributions, it turns out that component U lies $\sim$10\,pc closer than A (both in LCC). Component C (USCO) has an average distance similar to G, but it is skewed towards larger distances.
Components D and F overlap in the sky, but F appears to be slightly more distant. 
Component E is double-peaked with the more distant peak overlapping with USCO in the sky. Stars from the peak at $\sim$135\,pc are distributed over the USCO and south of this region down to b$\approx$0\,deg.
Bin sizes are different for different components, in order make it possible to visualise components with very different total memberships on the same plot.
\href{https://github.com/mikeireland/chronostar/blob/master/projects/scocen/distances_components.py}{\faGithub}
}
\label{fig.distance_distribution_components}
\end{figure*}

Despite the partial overlap of components T and A in the sky,  \autoref{fig.distance_distribution_components} demonstrates that the majority of stars in A are located $\sim20$\,pc closer. This is a sparse component with a large velocity dispersion (1.8\,$\mathrm{km\,s^{-1}}$).

\begin{figure*}
\includegraphics[width=\linewidth]{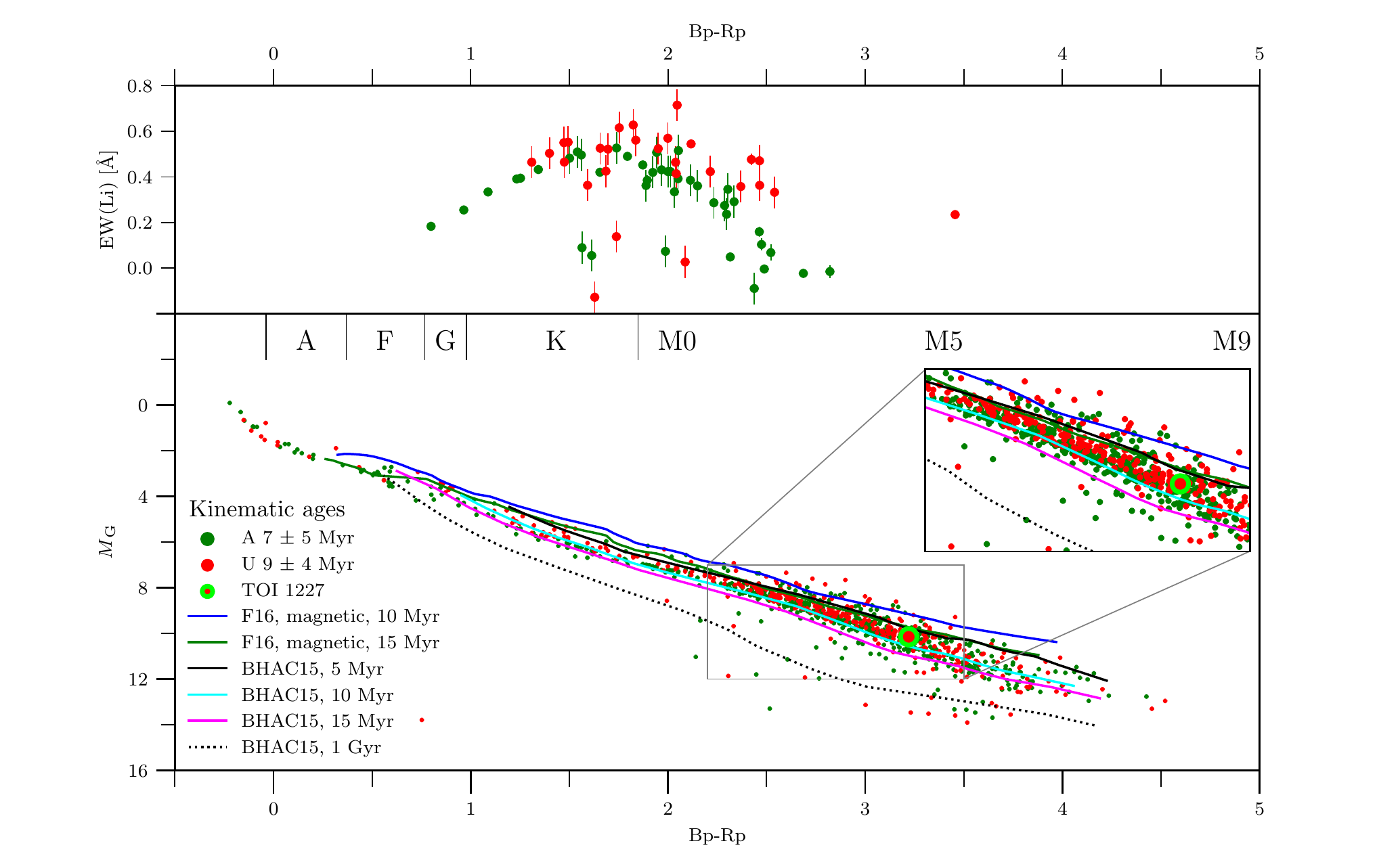}
\caption{\textit{Lower}: A Gaia $M_\text{G} - (B_\text{p}-R_\text{p})$ colour-magnitude diagram of our candidates in components A and U from Lower Centaurus-Crux with membership probability greater than 90\%. 
Their kinematic ages are the same within their uncertainties, but component U appears to be slightly younger than A based on its lithium strength. The inset shows the equal mass binary sequence in both components. An exoplanetary host TOI\,1227 belongs to component U.
\textit{Upper}: EW(Li) measurements for candidates as a function of Gaia $B_\text{p}-R_\text{p}$. 
\href{https://github.com/mikeireland/chronostar/blob/master/projects/scocen/cmd_age_sequence_AU.py}{\faGithub}
}
\label{fig.cmd_li_AU}
\end{figure*}

A colour-magnitude diagram shows the young age of both components and reveals that component U is likely younger than component A (\autoref{fig.cmd_li_AU}). Age is discussed in more detail in  \autoref{sec.age_validation}.

\subsection{Component C: Upper Scorpius}
The most studied part of the Sco-Cen association -- Upper Scorpius -- is described with the component C comprising 1,432 members. A cross-match with the list of 1,342 members from \citet{2020AJ....160...44L} that have \texttt{parallax\_error}$<$0.3\,mas shows 891 stars in common with C (membership probability $>$0.5), and additional 373 stars from components E and B.


This is a complex component in the \textsc{Chronostar} model, consistent with the growing observational evidence for multiple populations in USCO (e.g. \citealp{2021ApJ...917...23K,2021A&A...647A..14G,2021MNRAS.tmp.1862S}). It likely consists of multiple populations that have not been recovered within this model due to the limitations of the input catalog to the fit. This claim is supported by the relatively large velocity dispersion of 1.2\,$\mathrm{km\,s^{-1}}$ in each dimension (\autoref{tab.components_fit}) and a large luminosity spread of about 2 magnitudes for M0 dwarfs, which is a few times wider than for the T component (about 0.6\,mag; \autoref{fig.cmd_li_CUT}). This might result from component C containing multiple populations of different ages. On the other hand, the distance distribution for C does not reveal any distinct overdensities.


\subsection{Component E}
The diffuse part of the association between Upper Scorpius and Corona Australis is described by component E, which also overlaps significantly with the Upper Scorpius region and partly the Lupus complex in the sky. This is one of the complex and poorly-constrained components because the majority of its members lie outside of our input data to the fit. Its colour-magnitude diagram shows a double sequence where the younger sequence comes from the Lupus overdensity. Likewise, its distance distribution (\autoref{fig.distance_distribution_components}) shows two major peaks. Stars from the more distant peak lie slightly behind the southern part of USCO, while objects from the nearer peak are more evenly distributed in the sky. An overdensity at (l,b)$\approx$(0,7)\,deg contains 46 out of 48 stars from the group Theia\,67 identified by \citet{2019AJ....158..122K}.

\subsection{Components F and D: V1062 Sco}
One of the most compact components of the model, component D, corresponds to the group V1062\,Sco that was recently discovered by \citet{2018A&A...614A..81R}. The authors report 63 comoving stars with ages from less than 10 to about 25\,Myr. Our component D contains 33 members from their list, component F 6 members and component B 10 members.
However, we find that component D contains many more stars than \citet{2018A&A...614A..81R} study: 506 stars above the 50\% membership probability threshold.


\textsc{Chronostar} assigns component D a kinematic age of 2\,Myr, but the large associated uncertainty of $\pm$7\,Myr makes this estimate highly unreliable; the component is dense and likely gravitationally bound, invalidating the central assumption of \textsc{Chronostar}'s kinematic aging method. This concern is consistent with a direct estimate of boundedness: the virial mass sum of component D (neglecting factors of order unity) is $\sim$150\,M$_\odot$, significantly less than the total cluster mass. 
Moreover, the colour-magnitude diagram for this component (\autoref{fig.cmd_li_GDF}) demonstrates an overlap with the component T again suggesting that kinematic analysis underestimates its age. 
The median distance derived of the members is 176\,pc, which compares well with 175\,pc as found by \citet{2018A&A...614A..81R}. 

Component F, which is described as SC\,14 in \citet{2021ApJ...917...23K}, is located in the same line of sight as D, but is more diffuse with no clear peak in the distance distribution. Its stars are found between 160 and 200\,pc.
Despite the fact that the number of stars used in the fit was relatively low (16; \autoref{tab.components_fit}), its kinematic age 15$\pm$2\,Myr seems to roughly compare with that of component T and is consistent with the isochronal age (14.1$\pm$1.2\,Myr) from \citet{2021ApJ...917...23K}.

\subsection{Components H and I: IC\,2602 and Platais\,8}
In addition to V1062\,Sco, \textsc{Chronostar} also identifies two other known open clusters: components H (IC\,2602) and I (Platais\,8). A cross-check with \cite{2010MNRAS.409.1002D} revealed that 11 out of 17 stars in their list of the IC\,2602 members that have 100\% membership probability in component H, which contains a total of 819 members in our model.
Component I has 552 members, 6 of which are in common with \citet{2018ApJ...856...23G}'s list of 11 stars from Platais\,8 \citep{1998AJ....116.2423P}. The median distance of Platais\,8 in this work is 143\,pc.

As for the D component, we regard \textsc{Chronostar}'s kinematic ages for these components unreliable due to their high density and potential gravitational interactions; consistent with this concern, the ages we recover have very large uncertainties. IC\,2602 has a reported lithium depletion boundary age of 46$^{+6}_{-5}$\,Myr \citep{2010MNRAS.409.1002D} while the age estimate for Platais\,8 is 79\,Myr with a large uncertainty \citep{2018A&A...615A..12Y}.




\begin{figure*}
\includegraphics[width=\linewidth]{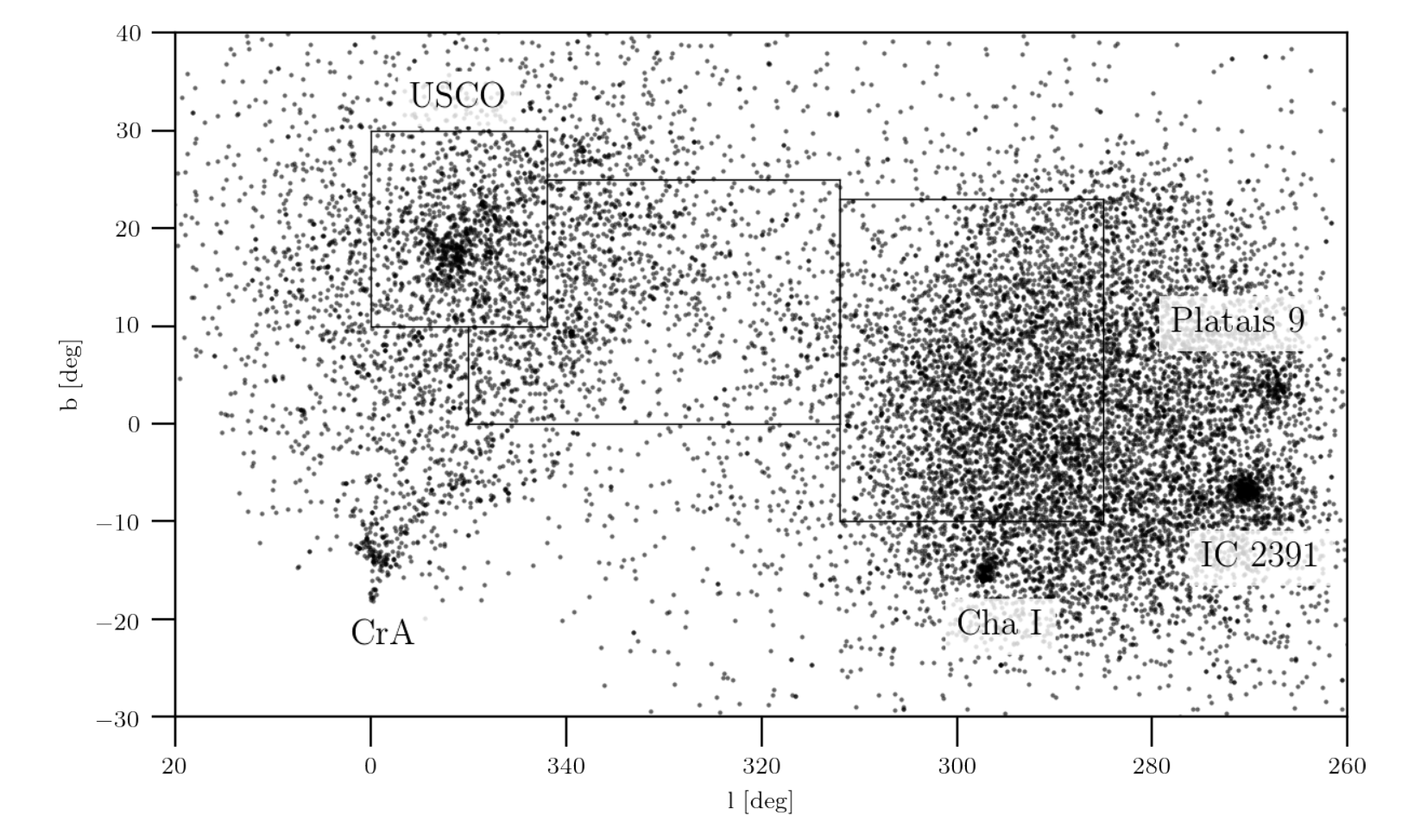}
\caption{Components B, J and Q from the model are complex and cover a very large volume. Out of their 15,432 
members, most are old stars and belong to the background. Due to their expanse, however, they also cover some of the overdense groups of stars, e.g. Corona Australis (CrA), Chamaeleontis I (Cha\,I),   
Upper Scorpius (USCO), IC\,2391 and Platais\,9.
\href{https://github.com/mikeireland/chronostar/tree/master/projects/scocen/galaxy_components_JBQ_paper.py}{\faGithub}
}
\label{fig.gx_black}
\end{figure*}

\subsection{Background and components B, J and Q} \label{sec.background}
Although \textsc{Chronostar}'s model incorporates a component that is associated with the background from the very beginning, its component splitting method introduced 6 additional components that typically consist of thousands of stars but completely lack young stars in the colour-magnitude diagram. They have large velocity dispersions and a very young age; the latter is required because it is impossible for such a kinematically incoherent group of stars to stay together for a long time. We denote these as background components (see  \autoref{tab.components_fit}). The most likely explanation for the introduction of this components is that the stellar background (i.e. stars not kinematically associated with the moving group) is not as homogeneous as \textsc{Chronostar} assumes, and that the fixed amplitude for \textsc{Chronostar}'s background is not high enough.

There are three dominant background components in the model: B, J and Q, consisting of 15,432 
members in total. They are centered on the outskirts of the association and are not well constrained due to the poor coverage of the parameter space outside the traditional Sco-Cen boundaries in the input catalog. Most of these stars are old background objects, but \autoref{fig.gx_black} reveals a substructure that corresponds to some of the known groups of young stars. Their colour-magnitude diagrams confirm the presence of a very young population of stars above the main sequence. These overdensities might be part of the known groups or might represent standalone components that would be recovered with a more populous input catalog. At the same time, it is likely that the large background components stole some of the members of the more diffuse components of the Sco-Cen association. Further fitting would help to answer these questions.

A further cross-check with the literature confirms membership of 299 of 313 stars reported to be in the Corona Australis association by \citet{2020A&A...634A..98G}. These authors found that the association appears to be more complex and is composed of two kinematically and spatially distinct subgroups with stars younger than 10\,Myr. Additionally, our cross-check with the list from \citet{2018ApJ...856...23G} confirms that our component B contains 7 X-ray sources from \citet{2000A&AS..146..323N} that lie in between Corona Australis and Upper Scorpius, and belong to the Upper Corona Australis in the Banyan\,$\Sigma$ model.

Other known young clusters and associations covered by the background components are IC\,2391, Platais\,9, the $\epsilon$\,Chamaeleontis star-forming complex \citep{2013MNRAS.435.1325M} that is split into Cha\,I and Cha\,II in \citep{2021A&A...646A..46G}. 
Component Q consists of 167 stars belonging to Cha\,I and 33 found in Cha\,II in their list. This is a very young complex, with an age estimated to be 1-2\,Myr \citep{2021A&A...646A..46G}.

\section{Discussion}\label{sec.age_validation}

\subsection{Kinematic ages versus ages derived by other methods}

Kinematic ages are based on the assumption that associations are gravitationally unbound and thus expanding. Stars in an association share their common motion through space and follow ballistic orbits around the Galaxy. The model in \textsc{Chronostar} uses this picture to assign ages, by searching for an initial distribution (in 6D phase space) and age such that, when the distribution is propagated forward by the modelled age of the component, it matches the 6D phase space distribution for that component which we observe today. 

Given the unavoidable assumptions behind this technique, it is important to cross-check kinematic ages against those derived from other methods, which carry their own assumptions and uncertainties. One cross-check to which we have already alluded is to ages derived from color-magnitude diagrams. \autoref{fig.cmd_li_CUT} compares color-magnitude diagrams of three components with a range of kinematic ages: component T (15$\pm$3\,Myr), U (9$\pm$4\,Myr) and C (4$\pm$4\,Myr; this is the youngest component (USCO) based on the isochronal fits from the literature, but it is non-trivial to interpret; see \autoref{sec.age_usco}). The trend of decreasing overluminosity with kinematic age is clear, and is also consistent with the relative difference in lithium strength between components T and C. Thus we find a comforting qualitative consistency between the relative ages that \textsc{Chronostar} derives from kinematics and the relative ages one would infer from both color-magnitude diagrams and lithium depletion. 

To make this comparison more quantitative, in \autoref{fig.cmd_li_CUT} we overplot \citet{2015A&A...577A..42B} isochrones for 5, 10 and 15\,Myr, as well as magnetic isochrones from \citet{2016A&A...593A..99F}; we have converted these to Gaia photometry using the colour--temperature and luminosity--$\log{g}$ relations from \citet{2013ApJS..208....9P} based on an online version of their table\footnote{ \url{https://www.pas.rochester.edu/~emamajek/EEM_dwarf_UBVIJHK_colors_Teff.txt}, version 2021.03.02}. 
Besides the fact that these relations are provided for the main sequence stars while magnetic isochrones are computed for young ages, neither of the \citet{2015A&A...577A..42B} isochrones matches the G or K stars perfectly either.

In the remainder of this section we make more detailed comparisons between our kinematic ages, isochronal ages based on our own colour-magnitude diagrams, and  previously-published literature estimates. We focus here on the components with the most well-determined kinematic ages (G, T, A, U, and C), excluding fits based on very few stars (F), complex components for which a single kinematic age is dubious (B, E, J, and Q), and collections of stars that are likely bound and thus fail to satisfy the basic assumption of kinematic aging (D, H, and I).



\subsubsection{Age of UCL: Components T and G}
\textsc{Chronostar}'s kinematic age for component T is 15$\pm$3\,Myr. The model is fit on 76 stars with measured radial velocities; most are FGK or early M-type dwarfs. Based on the colour-magnitude diagram shown in \autoref{fig.cmd_li_CUT}, this component has a relatively small spread in luminosity ($\sim$0.6\,mag for M0 dwarfs), likely indicating a small age spread. Comparison of the stars in this component to \citet{2015A&A...577A..42B} isochrones gives an age of 15-22\,Myr, consistent with the kinematic estimate. Both estimates are also consistent with the main-sequence turn-off age of 17$\pm$1\,Myr obtained by \citet{2002AJ....124.1670M}. 
Thus our result represents the first kinematic age derived for the the Upper\,Centaurus-Lupus association that agrees with age estimates from stellar evolution.

Our kinematic age estimate for component G is $13\pm 8$ Myr. Thus our central estimate of the age of this component is similar to that for component T, which is consistent with our finding that the two components overlap substantially in the colour-magnitude diagram. However, component G also shows a small dispersion in luminosity along the sequence (comparable to component T; \autoref{fig.cmd_li_GDF}), suggesting an age spread smaller than our $\pm 8$ Myr. Literature estimates confirm this intuition. For example, \citet{2002AJ....124.1670M} find intrinsic 1$\sigma$ age dispersions of 3\,Myr in UCL. The relatively large kinematic age spread that we obtain may be artificially inflated by the fact that the component is composed of multiple clumps (\autoref{fig.gx}) that, given a larger and more complete sample of kinematic data, would result in this component being subdivided further.

\subsubsection{Age of USCO: Component C} \label{sec.age_usco}
In contrast to the well constrained isochronal age of the Upper Centaurus-Lupus, literature estimates for the Upper Scorpius vary from 4-5\,Myr for low-mass stars \citep{2002AJ....124..404P} to 11-13\,Myr for intermediate-mass stars \citep{2017ApJ...842..123F, 2012ApJ...746..154P}. Similar age discrepancies have been found by \citet{2016ApJ...817..164R}. 
\citet{2016A&A...593A..99F} and \citet{2019ApJ...872..161D} found that these discrepancies could be due to magnetic inhibition of convection and underestimated masses for low-mass stars, and argue that taking these effects into account produces an age estimate of 10\,Myr for USCO, closer to the published value for intermediate stars than for low-mass stars.
\citet{2021ApJ...912..137S} also argue that a population of undetected binary stars could be responsible for the inconsistent age estimates derived for low- versus intermediate-mass stars, and that correcting for this effect yields a consistent age of $\sim$10\,Myr with a small intrinsic age spread \citep{2021ApJ...912..137S}. By contrast, \citet{2018MNRAS.476..381W} argue that USCO is an outcome of multiple bursts of star formation and thus has a real, substantial age spread. Similarly, \cite{2021MNRAS.tmp.1862S} claim that star formation in USCO must have lasted more than 10\,Myr.

Given that parts of component E overlap with component C, and some of the young stars in the USCO region are also members of component B, our model supports the view that Upper Scorpius is indeed a heterogeneous mixture of multiple populations with distinct kinematics and ages. Our kinematic age for component C is 4$\pm$4\,Myr, indicating an age spread comparable to the mean age. The large spread luminosity (about 2 magnitudes for M0 stars, a factor of a few larger than for component T; \autoref{fig.cmd_li_CUT}) supports the notion of multiple populations with different ages. More detailed kinematic analysis would be necessary to reliably disentangle the subcomponents and evaluate their individual ages.

\subsubsection{Age of LCC: Components A and U}
The isochronal age for the LCC has also been controversial, but to a lesser extent. 
As \citet{2016MNRAS.461..794P} point out, the ages of UCL and LCC have typically been found to be very similar, i.e. 16$\pm$1\,Myr for UCL and 17$\pm$1\,Myr for LCC \citep{2012ApJ...746..154P}. \citet{2002AJ....124.1670M}, however, find that LCC might be slightly younger than UCL, depending on the member selection. Moreover, \cite{1989A&A...216...44D} argue that UCL is sightly older, 14-15\,Myr, while LCC is slightly younger, 11-12\,Myr. 

LCC in our kinematic model is split into two parts: component A in the upper and component U in the lower section. The kinematic age of the component A is 7$\pm$5\,Myr, while that of component U is 9$\pm$4\,Myr. While component A overlaps with the component T in the CMD, component U does not: it appears to be more overluminous than component A, and also has higher lithium content (\autoref{fig.cmd_li_AU}).
This confirms a report from \citet{2008hsf2.book..235P} who state that ``there is some hint of substructure in the group, and it appears that the northern part of the group is somewhat more distant, older, and richer ($\sim$17\,Myr, 120\,pc) than the southern part of the group ($\sim$12\,Myr, 110\,pc).'' Similarly, the work of \citet{2018ApJ...868...32G} with ages based on the CIFIST evolutionary models \citep{2015A&A...577A..42B} concludes that their groups A0 and A (lower LCC) are somewhat younger, age $\sim$7\,Myr, than their groups B and C (upper LCC), with an age of 9-10\,Myr.
Likewise, \citet{2021ApJ...917...23K} report on age gradient in the LCC region decreasing from the north to the south. However, their northernmost part is older (23$\pm$2.3\,Myr) than our estimate, but the age of their component at $b=-10$\,deg (13$\pm$1.4\,Myr) that falls within our component U agrees with our kinematic age.

\subsubsection{Age of PDS\,70} \label{sec.pds70}
One of the members of component T is the T\,Tauri K star PDS\,70, which hosts the protoplanet PDS\,70b within a gap in its transition disk \citep{2018A&A...617A..44K, 2018A&A...617L...2M}. This is the first planet conclusively imaged during formation. \citet{2019NatAs...3..749H} found a second protoplanet, PDS\,70c, in the system, and \citet{Benisty_2021} recently discovered a dust-depleted cavity in a circumplanetary disk around this planet.


Several surveys have measured the radial velocity of this star. Gaia\,DR2 \citep{2018A&A...616A...6S,2018A&A...616A...1G} reports 3.1$\pm$1.4\,$\mathrm{km\,s^{-1}}$ while \citet{2020ApJ...892...81T} and \citet{2021MNRAS.503..938Z} measure 6.0$\pm$1.5\,$\mathrm{km\,s^{-1}}$ and 5.1$\pm$1.5\,$\mathrm{km\,s^{-1}}$, respectively. 
Using the Gaia radial velocity, as we do in our default analysis since this is the value with the smallest reported uncertainty, yields a 95\% membership probability in component T for this star. However, this does not change substantially if we use \citeauthor{2021MNRAS.503..938Z}'s value of 5.1$\pm$1.5\,$\mathrm{km\,s^{-1}}$ instead: the membership probability in this case is 90\%.
The location of the star in association is shown in \autoref{fig.gx}.


\citet{2018A&A...617L...2M} estimate the age of PDS\,70 to be 5.4\,$\pm$\,1.0\,Myr based on non-magnetic MIST models \citep{2016ApJ...823..102C} coupled with a prior probability distribution that is flat in age and follows a \citet{2003PASP..115..763C} initial mass function in mass.
This was already a very young age, largely inconsistent with the $>$10\,Myr consensus age  for stars in Upper Centaurus Lupus 
\citep{2012ApJ...746..154P}.
Our kinematic age for component T is 15$\pm$3\,Myr, which 
is in even greater tension with \citeauthor{2018A&A...617L...2M}'s age.
\autoref{fig.cmd_li_CUT} shows a narrow sequence of stars in the component, and constrains the age spread of stellar siblings to a small value. Comparison of the colour-magnitude diagram location of PDS\,70 with that of component C also indicates that PDS\,70 is likely older than Upper Scorpius.
Finally, PDS\,70's equivalent width of lithium (0.52$\pm$0.07\,\AA, \citealp{2021MNRAS.503..938Z}) places it in between the 8 and 10 Myr lithium isochrones. We therefore conclude that PDS\,70 is likely substantially older than \citeauthor{2018A&A...617L...2M}'s estimate.

\subsubsection{Age of TOI\,1227}
The southernmost part of the Lower Centaurus Crux (\autoref{fig.gx}) is home to an M5 dwarf star TOI\,1227 
that hosts a sub-Jovian young planet with an estimated mass of 0.5\,M$_\text{J}$ and a radius of 0.85$\pm$0.05\,R$_\text{J}$ (9.5\,R$_\text{\earth}$; 
\citealp{2021arXiv211009531M}). The authors suggest that young planet TOI\,1227\,b is still contracting and will evolve to an object with a radius smaller than 5\,R$_\text{\earth}$.
They use lithium, stellar rotation and the colour-magnitude diagram 
to derive an age of 11$\pm$2\,Myr. Our analysis assigns TOI\,1227 to component U with $\approx 100\%$ membership probability, and \citeauthor{2021arXiv211009531M}'s age estimate is consistent with our kinematic age for component U (9$\pm$4\,Myr). We do not have radial velocity for this object in our database, but if we recompute the membership probability using the radial velocity of 13.3$\pm$0.3\,km\,s$^{-1}$ reported by \citeauthor{2021arXiv211009531M}, we reaffirm its membership. 

\subsection{On the purity and completeness of our membership list}


The contamination of our membership list with old field stars is small, and thus the purity of our sample is very high.
Only individual stars are found on or even below the main sequence, most often in the low-mass region due to the missing radial velocities for M dwarfs and lower quality of astrometric measurements due to their intrinsically low luminosity. 
The tests based on the synthetic data with uncertainties typical of Gaia in \citetalias{2019MNRAS.489.3625C} show that \textsc{Chronostar} falsely assigns memberships in less than 1\%.

However, our membership list is only partly complete: our model for Sco-Cen, meaning the 8 well-defined components we identify, contains 
a total of 8,185 stars with membership probability greater than 0.5 (4,511 stars with 90\% membership probability). 
By contrast, the expected population of Sco-Cen is larger. \citet{2011MNRAS.416.3108R} estimate that it contains 10,000 members,
while \citet{2019A&A...623A.112D} report nearly 11,000 pre-main-sequence stars with less than 3\% field-star contamination and $\sim$3600 main-sequence candidate members with a larger (10-30\%) field-star contamination rate. Out of 8,185 members indentified from our work, 6,990 objects are in common with \citet{2019A&A...623A.112D}: 
1,584 stars in G, 1,290 in C, 1,146 in A, 967 in T, 932 in E, 461 in D, 378 in U and 232 in component F.

The reason for the lower numbers in our model is partly the higher purity of our method, which is less subject to field star contamination than that of \citet{2019A&A...623A.112D}; their selection is based on proper motions, transverse velocities, parallax and cuts in the color-absolute magnitude diagram, which is less constraining than our use of full 6D phase space coordinates.
However, our sample may also be smaller for other reasons. Our input catalogue is defined by our selection boxes on the sky, which may not be large enough to include all Sco-Cen members. Our analysis may also assign some Sco-Cen members to the complex, poorly-defined components B, J, and Q.

\begin{figure*}
\includegraphics[width=\linewidth]{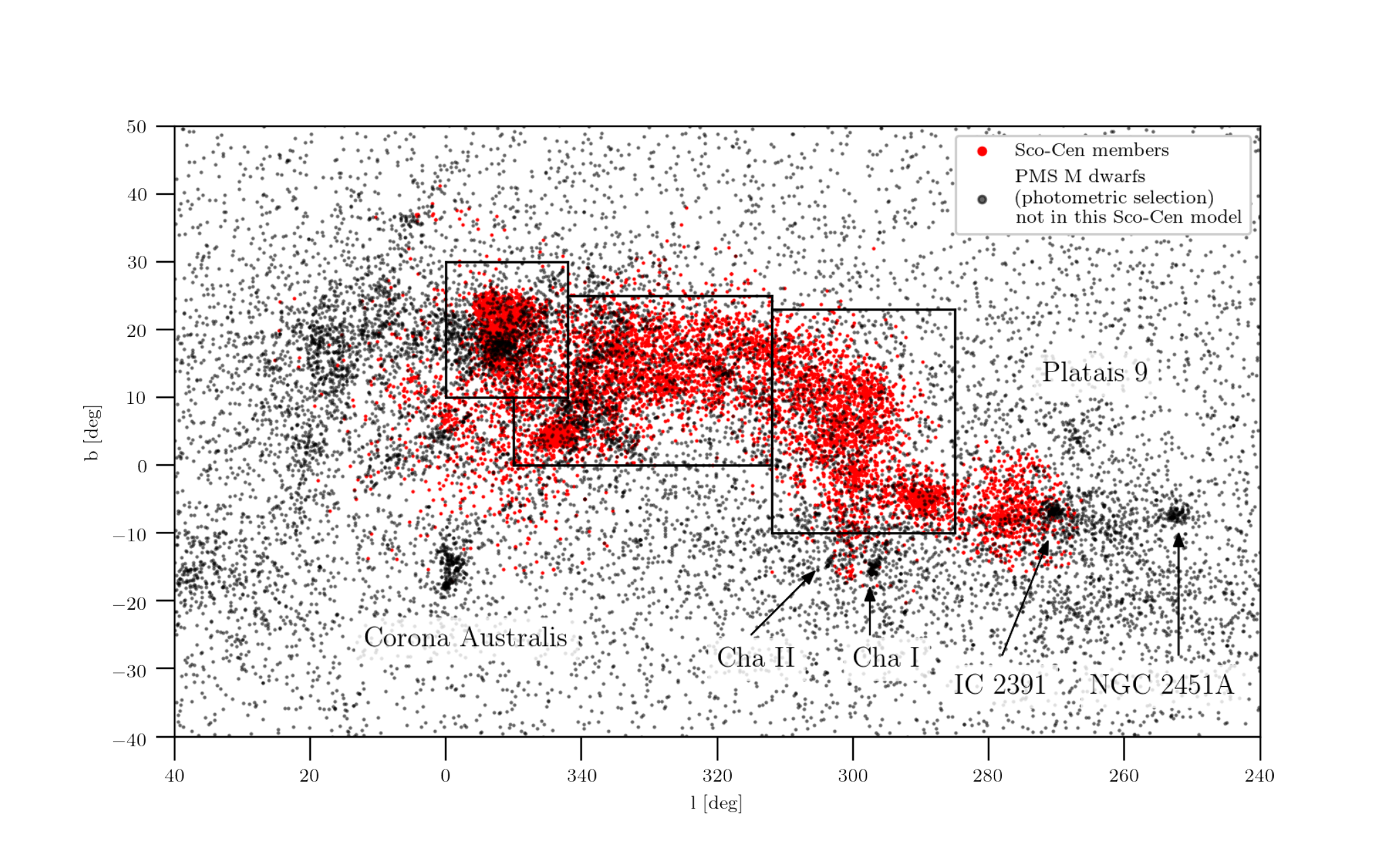}
\caption{
Sky positions of Sco-Cen members or potential members. Red points indicate stars kinematically-identified as belonging to Sco-Cen in this work, while black points show pre-main sequence M dwarfs at distances of 80 - 200 pc, selected following \citet{2021MNRAS.503..938Z}, that are \textit{not} included in our \textsc{Chronostar}-based membership list.
The PMS M dwarfs show
additional substructure both
within (e.g. in USCO) and around Sco-Cen
that is not captured by our kinematic analysis.
\href{https://github.com/mikeireland/chronostar/tree/master/projects/scocen/galaxy_black_with_field_PMS.py}{\faGithub} 
}
\label{fig.gx_field_PMS}
\end{figure*}

To understand these effects, we prepare
an alternative list of young star candidates
using photometric and distance selection; this will yield much lower purity than our kinematic catalog, but higher completeness, allowing us to identify places where our kinematic catalogue misses potential members. For this purpose, we follow the selection criteria from \citet{2021MNRAS.503..938Z} who found that in the sample of M dwarfs with $B_\text{P}-R_\text{P}>1.8$, \texttt{ruwe}\,$<1.4$ and $M_\mathrm{G}$ more than 1\,mag above the main sequence, 80\% of them show spectroscopic signs of youth, e.g.~lithium detection. With this selection we find 15,000 
candidate young stars 
between 80 and 200\,pc 
that are not members of our kinematic model but show overdensities within and around the traditional Sco-Cen boundaries in the sky; we show these stars, and our kinematically-selected Sco-Cen members for comparison, in \autoref{fig.gx_field_PMS}. We see that the most notable overdensities in our broader selection overlap with known clusters and associations, such as Corona Australis, 
Chamaeleon I and II, 
IC\,2391 and NGC\,2451A. The first three are known to be part of the Sco-Cen association but are not in our kinematic model either because they belong to the big complex components J, B or Q (e.g., Corona Australis, \autoref{fig.gx_black}) or because they were not in the input catalog to the fit (e.g. IC\,2391). It is clear that a more careful preparation of an initial catalog would result in a better model describing all significant parts of an association that would yield a higher completeness ratio of the membership list.

\section{Conclusions} \label{sec.conclusions}
We use \textsc{Chronostar} \citep{2019MNRAS.489.3625C} -- a robust Bayesian method for kinematic age determination of young stellar associations -- and Gaia\,DR2 data to construct a 6-dimensional kinematic model of the Sco-Cen association. Membership selection relies purely on the kinematics and completely neglects stellar age, location in the colour-magnitude diagram, or any other youth indicators.
The main conclusions of this work are:

\setlist[enumerate]{label={(\arabic*)}}
\begin{enumerate}
    \item The Sco-Cen model identifies 8 kinematically distinct components comprising more than 8,000 stars distributed in dense and diffuse groups. Upper Centaurus-Lupus and Lower Centaurus-Crux are split into two parts.
    \item Each of the components has an independently-fit kinematic age that is consistent with the isochronal age. This is the first time the kinematic age in a complex association is determined reliably.
    The kinematic age of the two Upper Cantaurus-Lupus components is 15\,$\pm$\,3\,Myr and 13\,$\pm$\,8\,Myr. Lower Centaurus-Crux, with ages of 7\,$\pm$\,5\,Myr and 9\,$\pm$\,4\,Myr for its two components, appears to be kinematically younger.
    \item We find that exoplanet host star PDS\,70 has a 97\% membership probability in component T of Upper Centaurus-Lupus. The kinematic age of this component is 15$\pm$3\,Myr; this narrow age range is supported by a small luminosity spread for the stars that make up this component.
\end{enumerate}

We conclude by noting that, while in this paper we have focused on Sco-Cen, there are obvious opportunities to our methods to other complex associations, which we intend to pursue in the future. Our analysis using \textsc{Chronostar} represents the first kinematic ages that reliably agree with age estimates from other techniques. However, our method has the great advantage that it does not rely on predefined, uncertain membership lists, and, as we demonstrate here, can be run blind or semi-blind to discover structures and assign ages to them. It is fundamentally scalable, in a way that isochronal or lithium depletion ages are not. Application of our method will give us access to a map of how star formation occurs in space and time with unprecedented detail. Current work in our group is additionally exploring non-spherical component models and the addition of isochronal ages to improve membership probabilities and enable finer substructure to be obtained.

\section*{Acknowledgements}
We acknowledge the traditional custodians of the land we work on, the Ngunnawal and Ngambri peoples, and pay our respects to elders past and present.

We thank Ronan Kerr for valuable suggestions that helped to improve this manuscript.

M{\v Z} acknowledges funding from the Australian Research Council (grant DP170102233) and support from the Consejer{\'{\i}}a de Econom{\'{\i}}a, Conocimiento y Empleo del Gobierno de Canarias and the European Regional Development Fund (ERDF) under grant with reference PROID2020010052.

This work has made use of data from the European Space Agency (ESA) mission {\it Gaia} (\url{https://www.cosmos.esa.int/gaia}), processed by the {\it Gaia} Data Processing and Analysis Consortium (DPAC,
\url{https://www.cosmos.esa.int/web/gaia/dpac/consortium}). Funding for the DPAC has been provided by national institutions, in particular the institutions participating in the {\it Gaia} Multilateral Agreement.

Software: \textsc{Chronostar} \citep{2019MNRAS.489.3625C}, \texttt{galpy} \citep{2015ApJS..216...29B}, \texttt{dustmaps} \citep{Green2018}, \texttt{astropy} \citep{astropy:2018}, \texttt{numpy} \citep{harris2020array}, \texttt{ipython} \citep{doi:10.1109/MCSE.2007.53} and \texttt{matplotlib} \citep{doi:10.1109/MCSE.2007.55}.


\section*{Data Availability}
Results from this work are provided in \autoref{tab.components_fit}, \autoref{tab.components_overlaps} and \autoref{tab.results}. Full \autoref{tab.components_fit} and \autoref{tab.results} are available online as supplementary material.
Gaia data \citep{2018A&A...616A...1G} is available on \href{https://gea.esac.esa.int/archive/}{https://gea.esac.esa.int/archive/}.
Radial velocity measurements are taken from the GALAH\,DR3 survey \citep{2021MNRAS.506..150B}, RAVE\,DR6 list \citep{2020AJ....160...82S}, APOGEE\,DR16 list \citep{2020ApJS..249....3A}, \citet{2021MNRAS.503..938Z}, and a compiled table in the Banyan\,$\Sigma$ paper \citep{2018ApJ...856...23G} - all references from their table are listed below \autoref{tab.results}.
Equivalent widths of lithium are taken from \citet{2021ApJ...908..247W}, \citet{2015MNRAS.448.2737R}, \citet{2019MNRAS.484.4591Z} and \citet{2021MNRAS.503..938Z}.
We used evolutionary models computed by \citet{2015A&A...577A..42B} for Gaia filters $G$, $B_\text{P}$  and $R_\text{P}$ and available on \url{http://perso.ens-lyon.fr/isabelle.baraffe/BHAC15dir/BHAC15_iso.GAIA}. Magnetic models from \citet{2016A&A...593A..99F} are found on \url{https://github.com/gfeiden/MagneticUpperSco/}. Empirical relations from "A Modern Mean Dwarf Stellar Color and Effective Temperature Sequence" \citep{2013ApJS..208....9P} are found on
\url{https://www.pas.rochester.edu/~emamajek/EEM_dwarf_UBVIJHK_colors_Teff.txt}.



\bibliographystyle{mnras}
\bibliography{paper} 




\appendix

\section{Epicyclic approximation} \label{sec.epicyclic}
Numerical integration of galactic orbits is replaced by a computationally efficient epicyclic approximation in the curvilinear coordinate system following \cite{2004MNRAS.352.1199M}. The origin of this non-inertial coordinate system is set in the local standard of rest (LSR) 
at the Galactocentric radius R{$_\mathrm{LSR}$} that rotates around the Galactic centre at a constant rate. 
The radial component $\xi$ points toward the Galactic center at all times. The tangential component $\eta$ 
is positive in the direction of the Galactic rotation and $\zeta$ is displacement from the Galactic plane. 
The position of a star ($\xi, \eta, \zeta$) at a given time $t$ close to the LSR is approximated with
\begin{equation}
\begin{aligned}
    \xi(t) &= \xi_0 + \dot{\xi}_0 \kappa^{-1} \; \sin{\kappa t} + (\dot{\eta}_0 - 2A\xi_0) \; (1 - \cos{\kappa t}) (2B)^{-1} \\
    \eta(t) &= \eta_0 - \dot{\xi}_0 \; (1 - \cos{\kappa t}) (2B)^{-1} + \dot{\eta}_0 \; (A \kappa t - (A-B)\sin{\kappa t}) (\kappa B)^{-1} \\
    &- \xi_0 \; 2A(A-B) \; (\kappa t - \sin{\kappa t}) (\kappa B)^{-1} \\
    \zeta(t) &= \zeta_0 \cos{\nu t} + \dot{\zeta}_0 \nu^{-1} \sin{\nu t}. \\
\end{aligned}
\label{eq.epi}
\end{equation}

Velocities are determined by the first derivatives of the position:
\begin{equation}
\begin{aligned}
    \dot{\xi}(t) &= \dot{\xi}_0 \cos{\kappa t} + (\dot{\eta}_0 - 2 A \xi_0) \kappa \sin{\kappa t} (2B)^{-1} \\
    \dot{\eta}(t) &= -\dot{\xi}_0 \kappa (2B)^{-1} \sin{\kappa t} + \dot{\eta}_0 B^{-1} (A-(A-B) \cos{\kappa t}) \\
    &- 2A\xi_0 (A-B) (1 -\cos{\kappa t}) B^{-1} \\
    \dot{\zeta}(t) &= -\zeta_0 \nu \sin{\nu t} + \dot{\zeta}_0 \cos{\nu t}.
\end{aligned}
\end{equation}

The initial positions ($\xi_0, \eta_0, \zeta_0$) and ($\dot{\xi}_0, \dot{\eta}_0, \dot{\zeta}_0$) are free parameters of the fit. In practice, \textsc{Chronostar} operates in the cartesian space. The conversion between the cartesian and curvilinear systems is in place solely for the epicyclic approximation. 

Epicyclic frequencies $\kappa$ and $\nu$ are related to the Oort's constants $A = 15.3 \pm 0.4 \, \mathrm{km \, s^{-1} \, kpc^{-1}}$ and $B = -11.9 \pm 0.4 \, \mathrm{km \,s^{-1} \, kpc^{-1}}$ \citep{2017MNRAS.468L..63B} by
\begin{equation}
\begin{aligned}
    \kappa &= \sqrt{-4 B (A - B)} \\
    \nu &= \sqrt{4 \pi G \rho + (A + B) (A - B)}
\end{aligned}
\end{equation}
where $\rho = 0.0889 \pm 0.0071 \; \mathrm{M_\odot \, pc^{-3}}$ \citep{2018PhRvL.121h1101S}.

\begin{figure}
\includegraphics[width=\linewidth]{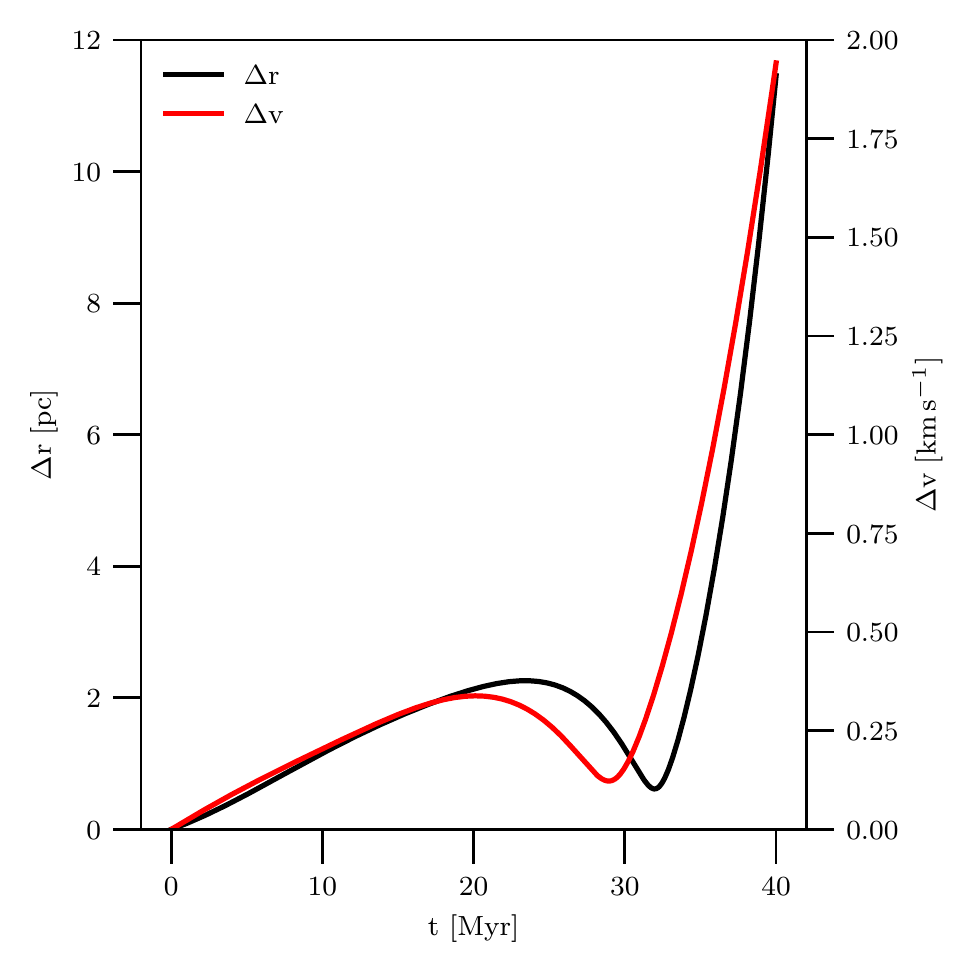}
\caption{Epicyclic orbit approximates numerically integrated \texttt{galpy} trajectory within $\Delta$r\,=\,2\,pc and $\Delta$v\,=\,0.3\,km\,s$^{-1}$ for $\sim$30\,Myr. $\Delta$r and $\Delta$v are absolute differences between \texttt{galpy} and epicyclic orbits.
\href{https://github.com/mikeireland/chronostar/tree/master/epicyclic_fine_tuning/test_makarov_in_traceorbit.py}{\faGithub}
}
\label{fig.epicyclic_vs_galpy}
\end{figure}

Since these values were determined from a much bigger volume of data, i.e. Oort's constants 
at a typical heliocentric distance of 230\,pc, and the density at the kiloparsec scale, we fine tuned constants A, B and $\rho$ to match the \texttt{galpy} orbits at the Sco-Cen distance more precisely:
\begin{equation}
A' = 0.89 \, A, \\
B' = 1.15 \, B, \\
\rho' = 1.21 \, \rho.
\end{equation}
This gives $\kappa = 0.04\,\mathrm{Myr^{-1}}$ 
and $\nu = 0.08\,\mathrm{Myr^{-1}}$. 
Such epicyclic approximation for a star in USCO 
is valid for $\sim$30\,Myr when the difference with \texttt{galpy} orbit integration starts to increase significantly over time, i.e. beyond 2\,pc and 0.3\,km\,s$^{-1}$ (\autoref{fig.epicyclic_vs_galpy}).

\section{Reddening correction} \label{sec.reddening}
Sco-Cen is located in the Galactic plane rich in dust. This means that the measured Gaia magnitudes in our catalogue are affected by interstellar extinction. To correct for reddening, we use the procedure from \citet{2021MNRAS.504.5788R}\footnote{\url{https://github.com/adrains/plumage}} that uses the 3D dust map of \citet{2020A&A...639A.138L} and is implemented in the python package \texttt{dustmaps} \citep{Green2018}. Extinction coefficients for Gaia\,DR2 $G$, $B_\text{P}$ and $R_\text{P}$ bands were determined from the $B_\text{P}-R_\text{P}$-dependent relation found in \citet{2021MNRAS.507.2684C}. We apply this relation on the entire sample but note that it is only valid for $0.2<(B_\text{P}-R_\text{P})_0<2$. 



\section{Membership probabilities} \label{sec.membership_probabilities}

\newcommand{\memberships}{\input{stellar_memberships_example.tex}}
\begin{landscape}
\begin{table*}
\hspace*{-8cm}
\begin{tabular}{c cc rc ccccccccccccccc cc cc cc}
source\_id & (BP-RP)$_0$ & G$_0$ & RV & $\sigma_\mathrm{RV}$ & X & Y & Z & U & V & W & $\sigma_\mathrm{X}$ & $\sigma_\mathrm{Y}$ & $\sigma_\mathrm{Z}$ & $\sigma_\mathrm{U}$ & $\sigma_\mathrm{V}$ & $\sigma_\mathrm{W}$ & comp & p & EW(Li) & $\sigma_\mathrm{EW(Li)}$ & Ref \\
\hline
Gaia\,DR2 & & & $\mathrm{km\,s^{-1}}$ & $\mathrm{km\,s^{-1}}$ & pc & pc & pc & $\mathrm{km\,s^{-1}}$ & $\mathrm{km\,s^{-1}}$ & $\mathrm{km\,s^{-1}}$ & pc & pc & pc & $\mathrm{km\,s^{-1}}$ & $\mathrm{km\,s^{-1}}$ & $\mathrm{km\,s^{-1}}$ & & & \AA & \AA & RV/Li \\
 \hline
 \input{stellar_memberships_example.tex}
\end{tabular}
\caption{List of stars that have membership probabilities at least 50\% in any of the components in the Scorpius-Centaurus association. Column \texttt{comp} denotes the component the star belongs to with the membership probability \texttt{p}.
(BP-RP)$_0$ and G$_0$ are corrected for extinction.
Reference column reports on the source for radial velocity/lithium.
References: 
(1) \citet{2019MNRAS.484.4591Z}, 
(2) \citet{2018A&A...616A...1G}, 
(3) \citet{1997A&A...328..187C}, 
(4) \citet{2009A&A...498..949M}, 
(5) \citet{2013A&A...551A..46L}, 
(6) \citet{2007ApJ...657..338P}, 
(7) \citet{2007AJ....133.2524W}, 
(8) \citet{2011AJ....141..187S}, 
(9) E. Bubar et al. 2018 in preparation, 
(10) \citet{2006AstL...32..759G}, 
(11) \citet{2020AJ....160...82S}, 
(12) \citet{2011ApJ...738..122C}, 
(13) \citet{2006A&A...460..695T}, 
(14) \citet{2021MNRAS.503..938Z}, 
(15) \citet{1953GCRV..C......0W}, 
(16) \citet{2007A&A...467.1147G}, 
(17) \citet{2012ApJ...745...56D}, 
(18) \citet{2013MNRAS.435.1325M}, 
(19) \citet{2020ApJS..249....3A}, 
(20) \citet{1988Ap&SS.148..163G}, 
(21) \citet{2021MNRAS.506..150B}, 
(22) \citet{2015MNRAS.448.2737R}, 
(23) \citet{2007AN....328..889K}, 
(24) \citet{1967IAUS...30...57E}.
Full table is available online as supplementary material. 
\href{https://github.com/mikeireland/chronostar/tree/master/projects/scocen/print_stellar_memberships_BIG_table_for_paper.py}{\faGithub}  \href{https://github.com/mikeireland/chronostar/blob/master/projects/scocen/print_stellar_memberships_BIG_table_for_paper_entire_fits_file_for_supplementary_material.py}{\faGithub}
}
\label{tab.results}
\end{table*}
\end{landscape}
\bsp	
\label{lastpage}
\end{document}

%% file: componentsfit.tex
A & 60 & -95 & 42 & 2.7 & -8.7 & 0.8 & 10 & 10 & 9 & 1.1 & 1.1 & 0.9 & 34 & -37 & 30 & 0.6 & -9.4 & 2.3 & 6 & 1.1 & 7 & 5 & 153 &  \\
B & 122 & -36 & 60 & 6.2 & -6.4 & 1.9 & 22 & 22 & 21 & 1.9 & 1.8 & 1.8 & 91 & -16 & 47 & 4.8 & -7.3 & 3.4 & 20 & 1.9 & 5 & 11 & 296 & bg* \\
C & 133 & -19 & 76 & 5.2 & -3.5 & -0.9 & 7 & 7 & 7 & 1.2 & 1.2 & 1.2 & 113 & -17 & 76 & 4.3 & -4.1 & 0.8 & 5 & 1.2 & 4 & 4 & 130 &  \\
D & 168 & -50 & 39 & 7.0 & -7.3 & 3.0 & 3 & 3 & 3 & 0.3 & 0.3 & 0.3 & 151 & -44 & 32 & 6.2 & -7.7 & 3.5 & 3 & 0.3 & 2 & 7 & 46 &  \\
E & 129 & -22 & 52 & 7.9 & -4.2 & 0.7 & 19 & 18 & 16 & 1.7 & 1.5 & 1.1 & 54 & 10 & 28 & 5.4 & -6.3 & 3.4 & 6 & 1.6 & 11 & 4 & 94 &  \\
F & 169 & -58 & 36 & 9.7 & -8.8 & 3.1 & 10 & 10 & 8 & 0.7 & 0.6 & 0.3 & 64 & 58 & -22 & 3.8 & -11.7 & 3.8 & 1 & 0.6 & 15 & 2 & 16 &  \\
G & 116 & -59 & 59 & 6.9 & -8.5 & 2.8 & 16 & 15 & 12 & 1.1 & 0.9 & 0.8 & 49 & 40 & 2 & 2.5 & -10.3 & 5.3 & 8 & 1.0 & 13 & 8 & 167 &  \\
H & 48 & -143 & 12 & 2.6 & -9.4 & 6.2 & 11 & 10 & 6 & 0.4 & 0.4 & 0.6 & 13 & 3 & -70 & -1.9 & -10.4 & 3.4 & 8 & 0.4 & 15 & 21 & 84 &  \\
I & 18 & -133 & 8 & 0.0 & -10.1 & 3.3 & 6 & 6 & 6 & 0.5 & 0.5 & 0.5 & 11 & -111 & 1 & -0.6 & -10.2 & 3.3 & 6 & 0.5 & 2 & 11 & 35 &  \\
J & 62 & -94 & 36 & 3.0 & -9.6 & 2.0 & 28 & 26 & 21 & 1.7 & 1.6 & 1.6 & 34 & -2 & 7 & -0.3 & -10.4 & 3.3 & 21 & 1.6 & 10 & 12 & 197 & bg* \\
K & 105 & -126 & 45 & -22.9 & -3.0 & -5.7 & 27 & 27 & 27 & 2.8 & 2.8 & 2.7 & 145 & -128 & 56 & -23.3 & -2.0 & -5.1 & 26 & 2.8 & 2 & 9 & 77 & bg \\
L & 89 & -184 & 31 & -21.2 & -5.6 & 5.8 & 37 & 34 & 26 & 2.1 & 1.9 & 2.1 & 289 & -209 & -38 & -24.1 & -0.2 & 5.5 & 28 & 1.9 & 11 & 14 & 37 & bg \\
M & 93 & -95 & 53 & -12.4 & -7.8 & 6.4 & 33 & 32 & 30 & 2.4 & 2.3 & 2.3 & 159 & -75 & 10 & -14.3 & -6.0 & 7.6 & 30 & 2.3 & 6 & 12 & 116 & bg \\
N & 49 & -90 & 39 & -18.9 & -2.1 & 0.7 & 22 & 22 & 22 & 4.7 & 4.7 & 4.7 & 51 & -90 & 39 & -18.9 & -2.1 & 0.7 & 22 & 4.7 & 0 & 5 & 245 & bg \\
O & 90 & -247 & 35 & -13.0 & -10.1 & 4.1 & 14 & 14 & 13 & 0.2 & 0.2 & 0.4 & 131 & -212 & 11 & -15.0 & -9.0 & 4.8 & 14 & 0.2 & 5 & 78 & 0 & bg \\
P & 58 & -228 & 43 & -15.0 & -4.4 & -1.1 & 22 & 22 & 22 & 4.9 & 4.9 & 4.9 & 62 & -227 & 43 & -15.1 & -4.3 & -1.0 & 22 & 4.9 & 0 & 4 & 1 & bg \\
Q & 59 & -156 & 35 & -13.1 & -3.3 & -0.1 & 24 & 24 & 23 & 4.9 & 4.9 & 4.9 & 70 & -154 & 35 & -13.3 & -3.0 & 0.2 & 23 & 4.9 & 1 & 5 & 212 & bg* \\
R & 123 & -99 & 41 & -23.4 & -1.6 & 3.6 & 25 & 25 & 25 & 2.6 & 2.6 & 2.6 & 127 & -99 & 40 & -23.5 & -1.5 & 3.7 & 25 & 2.6 & 0 & 10 & 93 & bg \\
S & 126 & -102 & 47 & -11.8 & -2.3 & 0.5 & 31 & 31 & 31 & 4.0 & 4.0 & 4.0 & 126 & -102 & 47 & -11.8 & -2.3 & 0.5 & 31 & 4.0 & 0 & 8 & 210 & bg \\
T & 98 & -87 & 64 & 4.9 & -7.8 & 1.5 & 17 & 16 & 13 & 1.1 & 0.9 & 0.5 & 43 & 28 & 5 & 0.2 & -9.3 & 5.0 & 3 & 1.0 & 15 & 3 & 128 &  \\
U & 54 & -91 & 19 & 1.4 & -7.6 & -0.8 & 7 & 7 & 6 & 0.7 & 0.7 & 0.5 & 38 & -26 & 21 & -1.1 & -8.0 & 0.3 & 3 & 0.7 & 9 & 4 & 54 &  \\

%% file: componentsfitbackground.tex
K & 105 & -126 & 45 & -22.9 & -3.0 & -5.7 & 27 & 27 & 27 & 2.8 & 2.8 & 2.7 & 145 & -128 & 56 & -23.3 & -2.0 & -5.1 & 26 & 2.8 & 2 & 9 & 77 \\
L & 89 & -184 & 31 & -21.2 & -5.6 & 5.8 & 37 & 34 & 26 & 2.1 & 1.9 & 2.1 & 289 & -209 & -38 & -24.1 & -0.2 & 5.5 & 28 & 1.9 & 11 & 14 & 37 \\
M & 93 & -95 & 53 & -12.4 & -7.8 & 6.4 & 33 & 32 & 30 & 2.4 & 2.3 & 2.3 & 159 & -75 & 10 & -14.3 & -6.0 & 7.6 & 30 & 2.3 & 6 & 12 & 116 \\
N & 49 & -90 & 39 & -18.9 & -2.1 & 0.7 & 22 & 22 & 22 & 4.7 & 4.7 & 4.7 & 51 & -90 & 39 & -18.9 & -2.1 & 0.7 & 22 & 4.7 & 0 & 5 & 245 \\
R & 123 & -99 & 41 & -23.4 & -1.6 & 3.6 & 25 & 25 & 25 & 2.6 & 2.6 & 2.6 & 127 & -99 & 40 & -23.5 & -1.5 & 3.7 & 25 & 2.6 & 0 & 10 & 93 \\
S & 126 & -102 & 47 & -11.8 & -2.3 & 0.5 & 31 & 31 & 31 & 4.0 & 4.0 & 4.0 & 126 & -102 & 47 & -11.8 & -2.3 & 0.5 & 31 & 4.0 & 0 & 8 & 210 \\

%% file: componentsfitcomplex.tex
B & 122 & -36 & 60 & 6.2 & -6.4 & 1.9 & 22 & 22 & 21 & 1.9 & 1.8 & 1.8 & 91 & -16 & 47 & 4.8 & -7.3 & 3.4 & 20 & 1.9 & 5 & 11 & 296  \\
C & 133 & -19 & 76 & 5.2 & -3.5 & -0.9 & 7 & 7 & 7 & 1.2 & 1.2 & 1.2 & 113 & -17 & 76 & 4.3 & -4.1 & 0.8 & 5 & 1.2 & 4 & 4 & 130   \\
E & 129 & -22 & 52 & 7.9 & -4.2 & 0.7 & 19 & 18 & 16 & 1.7 & 1.5 & 1.1 & 54 & 10 & 28 & 5.4 & -6.3 & 3.4 & 6 & 1.6 & 11 & 4 & 94   \\
G & 116 & -59 & 59 & 6.9 & -8.5 & 2.8 & 16 & 15 & 12 & 1.1 & 0.9 & 0.8 & 49 & 40 & 2 & 2.5 & -10.3 & 5.3 & 8 & 1.0 & 13 & 8 & 167 \\
J & 62 & -94 & 36 & 3.0 & -9.6 & 2.0 & 28 & 26 & 21 & 1.7 & 1.6 & 1.6 & 34 & -2 & 7 & -0.3 & -10.4 & 3.3 & 21 & 1.6 & 10 & 12 & 197 \\
Q & 59 & -156 & 35 & -13.1 & -3.3 & -0.1 & 24 & 24 & 23 & 4.9 & 4.9 & 4.9 & 70 & -154 & 35 & -13.3 & -3.0 & 0.2 & 23 & 4.9 & 1 & 5 & 212 \\

%% file: componentsfitgood.tex
A & 60 & -95 & 42 & 2.7 & -8.7 & 0.8 & 10 & 10 & 9 & 1.1 & 1.1 & 0.9 & 34 & -37 & 30 & 0.6 & -9.4 & 2.3 & 6 & 1.1 & 7 & 5 & 153   \\
T & 98 & -87 & 64 & 4.9 & -7.8 & 1.5 & 17 & 16 & 13 & 1.1 & 0.9 & 0.5 & 43 & 28 & 5 & 0.2 & -9.3 & 5.0 & 3 & 1.0 & 15 & 3 & 128   \\
U & 54 & -91 & 19 & 1.4 & -7.6 & -0.8 & 7 & 7 & 6 & 0.7 & 0.7 & 0.5 & 38 & -26 & 21 & -1.1 & -8.0 & 0.3 & 3 & 0.7 & 9 & 4 & 54   \\

%% file: componentsfitdense.tex
D & 168 & -50 & 39 & 7.0 & -7.3 & 3.0 & 3 & 3 & 3 & 0.3 & 0.3 & 0.3 & 151 & -44 & 32 & 6.2 & -7.7 & 3.5 & 3 & 0.3 & 2 & 7 & 46   \\
F & 169 & -58 & 36 & 9.7 & -8.8 & 3.1 & 10 & 10 & 8 & 0.7 & 0.6 & 0.3 & 64 & 58 & -22 & 3.8 & -11.7 & 3.8 & 1 & 0.6 & 15 & 2 & 16   \\
H & 48 & -143 & 12 & 2.6 & -9.4 & 6.2 & 11 & 10 & 6 & 0.4 & 0.4 & 0.6 & 13 & 3 & -70 & -1.9 & -10.4 & 3.4 & 8 & 0.4 & 15 & 21 & 84   \\
I & 18 & -133 & 8 & 0.0 & -10.1 & 3.3 & 6 & 6 & 6 & 0.5 & 0.5 & 0.5 & 11 & -111 & 1 & -0.6 & -10.2 & 3.3 & 6 & 0.5 & 2 & 11 & 35   \\

%% file: componentsoverlaps.tex
A & 1264 & 883 & 639 & 7 & 5  \\
B & 16984 & 12432 & 10226 & 5 & 11  \\
C & 1444 & 1253 & 1119 & 4 & 4  \\
D & 512 & 422 & 360 & 2 & 7  \\
E & 1926 & 1233 & 887 & 11 & 4  \\
F & 298 & 219 & 187 & 15 & 2  \\
G & 1713 & 1037 & 649 & 13 & 8  \\
T & 1056 & 755 & 586 & 15 & 3  \\
U & 490 & 410 & 374 & 9 & 4  \\

%% file: stellar_memberships_example.tex
6110141563309613056 & 1.63 & 11.49 & 3.1 & 1.4 & 79.7 & -71.6 & 62.3 & 2.3 & -5.6 & 1.4 & 0.4 & 0.3 & 0.2 & 1.0 & 0.9 & 0.5 & T & 0.95 & 0.52 & 0.07 & 2/14  \\
6074563978391541504 & 2.15 & 12.89 & 12.4 & 1.2 & 59.1 & -99.1 & 42.5 & 1.8 & -9.1 & 0.9 & 0.4 & 0.7 & 0.1 & 0.6 & 1.1 & 0.2 & A & 0.98 & 0.36 & 0.07 & 2/14  \\
6075267425304250880 & 1.96 & 13.23 & 8.3 & 5.8 & 83.7 & -125.8 & 54.3 & 1.7 & -5.7 & 0.6 & 0.4 & 0.5 & 0.1 & 3.2 & 4.8 & 1.1 & T & 0.59 & 0.20 & 0.01 & 2/1  \\
5969266120643392256 & 1.21 & 11.42 & 1.8 & 1.1 & 170.9 & -53.8 & 38.9 & 6.9 & -8.1 & 3.8 & 2.2 & 0.7 & 0.2 & 1.1 & 0.4 & 0.2 & D & 1.00 & 0.13 & 0.07 & 2/14  \\
6017469427618958208 & 0.93 & 10.79 & 2.9 & 1.2 & 169.7 & -52.3 & 42.5 & 8.4 & -7.6 & 3.2 & 1.5 & 0.5 & 0.2 & 1.1 & 0.4 & 0.2 & D & 0.93 & 0.40 & 0.07 & 2/14  \\
6106733691438736128 & 1.65 & 12.18 & 4.8 & 0.8 & 103.9 & -72.7 & 75.7 & 6.2 & -7.3 & 2.1 & 0.7 & 0.5 & 0.4 & 0.6 & 0.4 & 0.3 & T & 0.99 & 0.48 & 0.07 & 2/14  \\
6249055271112519936 & 1.37 & 11.17 & -4.4 & 7.4 & 134.3 & -13.3 & 86.8 & 7.8 & -2.9 & 0.6 & 1.3 & 0.1 & 0.6 & 6.7 & 0.7 & 3.1 & C & 1.00 & 0.47 & 0.02 & 2/22  \\
6051498079848359296 & 2.11 & 11.95 & -14.4 & 5.3 & 140.5 & -9.2 & 68.9 & -2.0 & 0.0 & -1.9 & 1.7 & 0.1 & 0.5 & 5.1 & 0.4 & 1.6 & C & 0.97 & 0.73 & 0.02 & 2/22  \\
6235172592479759232 & 1.63 & 11.80 & -3.1 & 6.9 & 130.2 & -34.5 & 82.2 & 5.5 & -6.1 & 1.0 & 1.1 & 0.3 & 0.5 & 6.1 & 1.6 & 2.7 & C & 0.83 & 0.48 & 0.02 & 2/1  \\
6086395513764172416 & 2.15 & 13.04 & 10.1 & 3.8 & 70.2 & -97.7 & 60.2 & 3.5 & -8.2 & 1.5 & 0.5 & 0.6 & 0.2 & 2.1 & 3.0 & 1.1 & T & 0.63 & 0.30 & 0.02 & 2/1  \\
6014331906769737984 & 2.03 & 12.61 & -1.4 & 6.4 & 108.6 & -46.3 & 62.9 & 3.4 & -6.8 & 1.4 & 0.8 & 0.3 & 0.3 & 5.6 & 2.4 & 2.0 & G & 0.94 & 0.48 & 0.01 & 2/1  \\
6244337679032642688 & 1.79 & 12.18 & -8.0 & 1.6 & 131.4 & -13.8 & 74.4 & 3.6 & -2.0 & 0.2 & 0.8 & 0.1 & 0.3 & 1.5 & 0.2 & 0.6 & C & 1.00 & 0.60 & 0.02 & 2/22  \\
6013772083553731968 & 1.63 & 11.79 & 4.9 & 7.4 & 120.4 & -56.1 & 66.8 & 8.9 & -7.7 & 3.0 & 1.3 & 0.6 & 0.4 & 6.4 & 3.0 & 2.2 & G & 0.50 & 0.31 & 0.02 & 2/1  \\
6242626387962771456 & 1.81 & 12.10 & -6.0 & 0.9 & 131.3 & -18.4 & 74.3 & 5.8 & -4.0 & -1.7 & 1.1 & 0.1 & 0.4 & 0.8 & 0.2 & 0.3 & C & 1.00 & 0.52 & 0.01 & 2/1  \\
6107169755177631872 & 1.67 & 12.58 & 2.6 & 4.3 & 94.4 & -105.7 & 62.8 & 0.2 & -3.8 & -0.6 & 0.8 & 0.9 & 0.3 & 2.7 & 3.1 & 1.1 & T & 1.00 & 0.30 & 0.01 & 2/1  \\
6093133488520130176 & 1.38 & 11.94 & 8.3 & 1.7 & 122.2 & -107.3 & 63.7 & 5.4 & -8.9 & 2.4 & 1.0 & 0.9 & 0.3 & 1.3 & 1.1 & 0.4 & G & 0.61 & 0.47 & 0.07 & 2/14  \\
6099876213515229184 & 1.47 & 11.91 & 8.1 & 1.0 & 110.5 & -87.1 & 64.0 & 7.2 & -8.7 & 2.0 & 0.9 & 0.7 & 0.3 & 0.8 & 0.6 & 0.3 & T & 0.99 & 0.43 & 0.07 & 2/14  \\
6038520956430662144 & 1.91 & 12.46 & -2.7 & 2.1 & 124.0 & -30.0 & 59.9 & 6.1 & -4.6 & 0.4 & 0.6 & 0.1 & 0.2 & 2.0 & 0.5 & 0.6 & E & 0.96 & 0.55 & 0.03 & 2/22  \\
6208713089974550656 & 1.95 & 13.04 & -5.7 & 6.5 & 125.2 & -47.3 & 75.1 & 0.8 & -4.7 & 0.8 & 1.0 & 0.4 & 0.4 & 5.7 & 2.2 & 2.3 & G & 0.80 & 0.36 & 0.01 & 2/1  \\
6050978698041931776 & 1.75 & 11.89 & -8.1 & 9.9 & 131.2 & -13.7 & 67.1 & 4.2 & -2.6 & -2.8 & 0.7 & 0.1 & 0.2 & 9.4 & 1.0 & 3.0 & C & 1.00 & 0.55 & 0.01 & 2/1  \\
6039021371655978112 & 1.37 & 11.05 & -7.5 & 3.4 & 150.6 & -28.2 & 66.0 & 2.6 & -3.1 & -0.9 & 2.3 & 0.4 & 0.6 & 3.2 & 0.6 & 0.9 & E & 0.86 & 0.22 & 0.02 & 2/1  \\
5215814919990121856 & 1.77 & 11.51 & 17.0 & 1.4 & 38.0 & -98.2 & -10.2 & -1.0 & -8.1 & -2.9 & 0.4 & 1.1 & 0.4 & 0.5 & 1.3 & 0.5 & U & 0.71 & 0.48 & 0.07 & 2/14  \\
6107185251419688320 & 1.87 & 12.73 & 6.5 & 6.0 & 88.3 & -99.3 & 60.8 & 2.7 & -6.6 & 0.7 & 0.7 & 0.8 & 0.3 & 3.8 & 4.3 & 1.6 & T & 1.00 & 0.42 & 0.00 & 2/1  \\
6260746481328616448 & 1.83 & 12.52 & -5.9 & 3.9 & 132.9 & -15.2 & 93.9 & 6.1 & -3.5 & 0.5 & 1.1 & 0.1 & 0.6 & 3.4 & 0.4 & 1.8 & C & 0.94 & 0.57 & 0.01 & 2/1  \\
6135468641852927744 & 2.14 & 12.85 & 9.6 & 0.7 & 55.4 & -84.0 & 58.0 & 2.4 & -8.0 & 1.2 & 0.2 & 0.4 & 0.1 & 0.4 & 0.5 & 0.2 & T & 0.56 & 0.21 & 0.07 & 2/14  \\
6102544590795328896 & 1.98 & 12.85 & 3.5 & 2.5 & 122.5 & -100.2 & 70.1 & 2.1 & -7.6 & 0.4 & 4.7 & 3.8 & 1.7 & 1.9 & 1.6 & 0.7 & T & 0.94 & 0.38 & 0.01 & 2/1  \\
6049726075120913152 & 1.30 & 11.64 & -4.1 & 1.5 & 146.9 & -22.5 & 75.8 & 6.5 & -4.1 & 0.6 & 1.0 & 0.2 & 0.3 & 1.4 & 0.2 & 0.5 & C & 0.90 & 0.35 & 0.07 & 2/14  \\
6048935358761628288 & 1.84 & 11.90 & -8.0 & 6.7 & 130.5 & -17.7 & 63.3 & 3.4 & -2.0 & -1.9 & 0.7 & 0.1 & 0.2 & 6.4 & 0.9 & 1.9 & C & 0.99 & 0.52 & 0.02 & 2/22  \\
6050388913137070848 & 1.86 & 12.62 & -4.7 & 1.4 & 130.4 & -18.1 & 71.7 & 6.9 & -4.1 & -1.4 & 0.9 & 0.1 & 0.3 & 1.3 & 0.2 & 0.5 & C & 1.00 & 0.43 & 0.01 & 2/1  \\
6017410053991075584 & 1.14 & 11.41 & 0.9 & 1.1 & 165.4 & -50.4 & 40.1 & 6.5 & -7.4 & 3.1 & 1.4 & 0.4 & 0.1 & 1.0 & 0.4 & 0.1 & D & 1.00 & 0.54 & 0.07 & 2/14  \\
6043809637411589376 & 1.39 & 11.59 & -3.0 & 1.3 & 146.2 & -26.7 & 78.0 & 7.0 & -7.0 & 0.1 & 0.9 & 0.2 & 0.3 & 1.2 & 0.3 & 0.4 & C & 0.65 & 0.40 & 0.07 & 2/14  \\
6240832435964142080 & 1.25 & 11.30 & -3.2 & 0.5 & 125.7 & -23.7 & 84.8 & 6.5 & -5.1 & 2.4 & 0.7 & 0.1 & 0.3 & 0.5 & 0.1 & 0.2 & E & 0.54 & 0.39 & 0.07 & 2/14  \\
6237148384817472256 & 1.74 & 12.28 & -7.5 & 1.3 & 129.5 & -24.8 & 79.4 & 2.9 & -4.7 & -0.6 & 0.8 & 0.2 & 0.3 & 1.2 & 0.3 & 0.5 & C & 1.00 & 0.41 & 0.01 & 2/1  \\
6244019542218710912 & 1.52 & 11.31 & -17.1 & 4.9 & 125.5 & -16.4 & 78.6 & -4.0 & -2.4 & -5.7 & 0.9 & 0.1 & 0.4 & 4.5 & 0.6 & 1.9 & C & 1.00 & 0.59 & 0.02 & 2/22  \\
6127032325352993024 & 1.89 & 12.24 & 11.7 & 0.8 & 50.6 & -96.0 & 50.8 & 1.8 & -8.0 & 0.6 & 0.2 & 0.4 & 0.1 & 0.4 & 0.7 & 0.2 & A & 0.96 & 0.39 & 0.02 & 2/1  \\

%% file: paper.bbl
\begin{thebibliography}{}
\makeatletter
\relax
\def\mn@urlcharsother{\let\do\@makeother \do\$\do\&\do\#\do\^\do\_\do\%\do\~}
\def\mn@doi{\begingroup\mn@urlcharsother \@ifnextchar [ {\mn@doi@}
  {\mn@doi@[]}}
\def\mn@doi@[#1]#2{\def\@tempa{#1}\ifx\@tempa\@empty \href
  {http://dx.doi.org/#2} {doi:#2}\else \href {http://dx.doi.org/#2} {#1}\fi
  \endgroup}
\def\mn@eprint#1#2{\mn@eprint@#1:#2::\@nil}
\def\mn@eprint@arXiv#1{\href {http://arxiv.org/abs/#1} {{\tt arXiv:#1}}}
\def\mn@eprint@dblp#1{\href {http://dblp.uni-trier.de/rec/bibtex/#1.xml}
  {dblp:#1}}
\def\mn@eprint@#1:#2:#3:#4\@nil{\def\@tempa {#1}\def\@tempb {#2}\def\@tempc
  {#3}\ifx \@tempc \@empty \let \@tempc \@tempb \let \@tempb \@tempa \fi \ifx
  \@tempb \@empty \def\@tempb {arXiv}\fi \@ifundefined
  {mn@eprint@\@tempb}{\@tempb:\@tempc}{\expandafter \expandafter \csname
  mn@eprint@\@tempb\endcsname \expandafter{\@tempc}}}

\bibitem[\protect\citeauthoryear{{Ahumada} et~al.,}{{Ahumada}
  et~al.}{2020}]{2020ApJS..249....3A}
{Ahumada} R.,  et~al., 2020, \mn@doi [\apjs] {10.3847/1538-4365/ab929e}, \href
  {https://ui.adsabs.harvard.edu/abs/2020ApJS..249....3A} {249, 3}

\bibitem[\protect\citeauthoryear{{Baraffe}, {Homeier}, {Allard}  \&
  {Chabrier}}{{Baraffe} et~al.}{2015}]{2015A&A...577A..42B}
{Baraffe} I.,  {Homeier} D.,  {Allard} F.,   {Chabrier} G.,  2015, \mn@doi
  [\aap] {10.1051/0004-6361/201425481}, \href
  {https://ui.adsabs.harvard.edu/abs/2015A&A...577A..42B} {577, A42}

\bibitem[\protect\citeauthoryear{Benisty et~al.,}{Benisty
  et~al.}{2021}]{Benisty_2021}
Benisty M.,  et~al., 2021, \mn@doi [The Astrophysical Journal Letters]
  {10.3847/2041-8213/ac0f83}, 916, L2

\bibitem[\protect\citeauthoryear{{Bobylev} \& {Baykova}}{{Bobylev} \&
  {Baykova}}{2020}]{2020ARep...64..326B}
{Bobylev} V.~V.,  {Baykova} A.~T.,  2020, \mn@doi [Astronomy Reports]
  {10.1134/S1063772920040022}, \href
  {https://ui.adsabs.harvard.edu/abs/2020ARep...64..326B} {64, 326}

\bibitem[\protect\citeauthoryear{{Bovy}}{{Bovy}}{2015}]{2015ApJS..216...29B}
{Bovy} J.,  2015, \mn@doi [\apjs] {10.1088/0067-0049/216/2/29}, \href
  {https://ui.adsabs.harvard.edu/abs/2015ApJS..216...29B} {216, 29}

\bibitem[\protect\citeauthoryear{{Bovy}}{{Bovy}}{2017}]{2017MNRAS.468L..63B}
{Bovy} J.,  2017, \mn@doi [\mnras] {10.1093/mnrasl/slx027}, \href
  {https://ui.adsabs.harvard.edu/abs/2017MNRAS.468L..63B} {468, L63}

\bibitem[\protect\citeauthoryear{{Buder} et~al.,}{{Buder}
  et~al.}{2021}]{2021MNRAS.506..150B}
{Buder} S.,  et~al., 2021, \mn@doi [\mnras] {10.1093/mnras/stab1242}, \href
  {https://ui.adsabs.harvard.edu/abs/2021MNRAS.506..150B} {506, 150}

\bibitem[\protect\citeauthoryear{{Casagrande} et~al.,}{{Casagrande}
  et~al.}{2021}]{2021MNRAS.507.2684C}
{Casagrande} L.,  et~al., 2021, \mn@doi [\mnras] {10.1093/mnras/stab2304},
  \href {https://ui.adsabs.harvard.edu/abs/2021MNRAS.507.2684C} {507, 2684}

\bibitem[\protect\citeauthoryear{{Chabrier}}{{Chabrier}}{2003}]{2003PASP..115..763C}
{Chabrier} G.,  2003, \mn@doi [\pasp] {10.1086/376392}, \href
  {https://ui.adsabs.harvard.edu/abs/2003PASP..115..763C} {115, 763}

\bibitem[\protect\citeauthoryear{{Chen}, {Mamajek}, {Bitner}, {Pecaut}, {Su}
  \& {Weinberger}}{{Chen} et~al.}{2011}]{2011ApJ...738..122C}
{Chen} C.~H.,  {Mamajek} E.~E.,  {Bitner} M.~A.,  {Pecaut} M.,  {Su} K. Y.~L.,
   {Weinberger} A.~J.,  2011, \mn@doi [\apj] {10.1088/0004-637X/738/2/122},
  \href {https://ui.adsabs.harvard.edu/abs/2011ApJ...738..122C} {738, 122}

\bibitem[\protect\citeauthoryear{{Choi}, {Dotter}, {Conroy}, {Cantiello},
  {Paxton}  \& {Johnson}}{{Choi} et~al.}{2016}]{2016ApJ...823..102C}
{Choi} J.,  {Dotter} A.,  {Conroy} C.,  {Cantiello} M.,  {Paxton} B.,
  {Johnson} B.~D.,  2016, \mn@doi [\apj] {10.3847/0004-637X/823/2/102}, \href
  {https://ui.adsabs.harvard.edu/abs/2016ApJ...823..102C} {823, 102}

\bibitem[\protect\citeauthoryear{{Covino}, {Alcala}, {Allain}, {Bouvier},
  {Terranegra}  \& {Krautter}}{{Covino} et~al.}{1997}]{1997A&A...328..187C}
{Covino} E.,  {Alcala} J.~M.,  {Allain} S.,  {Bouvier} J.,  {Terranegra} L.,
  {Krautter} J.,  1997, \aap, \href
  {https://ui.adsabs.harvard.edu/abs/1997A&A...328..187C} {328, 187}

\bibitem[\protect\citeauthoryear{{Crundall}, {Ireland}, {Krumholz},
  {Federrath}, {{\v{Z}}erjal}  \& {Hansen}}{{Crundall}
  et~al.}{2019}]{2019MNRAS.489.3625C}
{Crundall} T.~D.,  {Ireland} M.~J.,  {Krumholz} M.~R.,  {Federrath} C.,
  {{\v{Z}}erjal} M.,   {Hansen} J.~T.,  2019, \mn@doi [\mnras]
  {10.1093/mnras/stz2376}, \href
  {https://ui.adsabs.harvard.edu/abs/2019MNRAS.489.3625C} {489, 3625}

\bibitem[\protect\citeauthoryear{{Dahm}, {Slesnick}  \& {White}}{{Dahm}
  et~al.}{2012}]{2012ApJ...745...56D}
{Dahm} S.~E.,  {Slesnick} C.~L.,   {White} R.~J.,  2012, \mn@doi [\apj]
  {10.1088/0004-637X/745/1/56}, \href
  {https://ui.adsabs.harvard.edu/abs/2012ApJ...745...56D} {745, 56}

\bibitem[\protect\citeauthoryear{{Damiani}, {Prisinzano}, {Pillitteri},
  {Micela}  \& {Sciortino}}{{Damiani} et~al.}{2019}]{2019A&A...623A.112D}
{Damiani} F.,  {Prisinzano} L.,  {Pillitteri} I.,  {Micela} G.,   {Sciortino}
  S.,  2019, \mn@doi [\aap] {10.1051/0004-6361/201833994}, \href
  {https://ui.adsabs.harvard.edu/abs/2019A&A...623A.112D} {623, A112}

\bibitem[\protect\citeauthoryear{{David}, {Hillenbrand}, {Gillen}, {Cody},
  {Howell}, {Isaacson}  \& {Livingston}}{{David}
  et~al.}{2019}]{2019ApJ...872..161D}
{David} T.~J.,  {Hillenbrand} L.~A.,  {Gillen} E.,  {Cody} A.~M.,  {Howell}
  S.~B.,  {Isaacson} H.~T.,   {Livingston} J.~H.,  2019, \mn@doi [\apj]
  {10.3847/1538-4357/aafe09}, \href
  {https://ui.adsabs.harvard.edu/abs/2019ApJ...872..161D} {872, 161}

\bibitem[\protect\citeauthoryear{{Dobbie}, {Lodieu}  \& {Sharp}}{{Dobbie}
  et~al.}{2010}]{2010MNRAS.409.1002D}
{Dobbie} P.~D.,  {Lodieu} N.,   {Sharp} R.~G.,  2010, \mn@doi [\mnras]
  {10.1111/j.1365-2966.2010.17355.x}, \href
  {https://ui.adsabs.harvard.edu/abs/2010MNRAS.409.1002D} {409, 1002}

\bibitem[\protect\citeauthoryear{{Evans}}{{Evans}}{1967}]{1967IAUS...30...57E}
{Evans} D.~S.,  1967, in {Batten} A.~H.,  {Heard} J.~F.,  eds, ~ Vol. 30,
  Determination of Radial Velocities and their Applications. p.~57

\bibitem[\protect\citeauthoryear{{Fang}, {Herczeg}  \& {Rizzuto}}{{Fang}
  et~al.}{2017}]{2017ApJ...842..123F}
{Fang} Q.,  {Herczeg} G.~J.,   {Rizzuto} A.,  2017, \mn@doi [\apj]
  {10.3847/1538-4357/aa74ca}, \href
  {https://ui.adsabs.harvard.edu/abs/2017ApJ...842..123F} {842, 123}

\bibitem[\protect\citeauthoryear{{Feiden}}{{Feiden}}{2016}]{2016A&A...593A..99F}
{Feiden} G.~A.,  2016, \mn@doi [\aap] {10.1051/0004-6361/201527613}, \href
  {https://ui.adsabs.harvard.edu/abs/2016A&A...593A..99F} {593, A99}

\bibitem[\protect\citeauthoryear{{Gagn{\'e}} et~al.,}{{Gagn{\'e}}
  et~al.}{2018}]{2018ApJ...856...23G}
{Gagn{\'e}} J.,  et~al., 2018, \mn@doi [\apj] {10.3847/1538-4357/aaae09}, \href
  {https://ui.adsabs.harvard.edu/abs/2018ApJ...856...23G} {856, 23}

\bibitem[\protect\citeauthoryear{{Gaia Collaboration} et~al.,}{{Gaia
  Collaboration} et~al.}{2016a}]{2016A&A...595A...1G}
{Gaia Collaboration} et~al., 2016a, \mn@doi [\aap]
  {10.1051/0004-6361/201629272}, \href
  {https://ui.adsabs.harvard.edu/abs/2016A&A...595A...1G} {595, A1}

\bibitem[\protect\citeauthoryear{{Gaia Collaboration} et~al.,}{{Gaia
  Collaboration} et~al.}{2016b}]{2016A&A...595A...2G}
{Gaia Collaboration} et~al., 2016b, \mn@doi [\aap]
  {10.1051/0004-6361/201629512}, \href
  {https://ui.adsabs.harvard.edu/abs/2016A&A...595A...2G} {595, A2}

\bibitem[\protect\citeauthoryear{{Gaia Collaboration} et~al.,}{{Gaia
  Collaboration} et~al.}{2018}]{2018A&A...616A...1G}
{Gaia Collaboration} et~al., 2018, \mn@doi [\aap]
  {10.1051/0004-6361/201833051}, \href
  {https://ui.adsabs.harvard.edu/abs/2018A&A...616A...1G} {616, A1}

\bibitem[\protect\citeauthoryear{{Gaia Collaboration} et~al.,}{{Gaia
  Collaboration} et~al.}{2021}]{2021A&A...649A...1G}
{Gaia Collaboration} et~al., 2021, \mn@doi [\aap]
  {10.1051/0004-6361/202039657}, \href
  {https://ui.adsabs.harvard.edu/abs/2021A&A...649A...1G} {649, A1}

\bibitem[\protect\citeauthoryear{{Galli}, {Bouy}, {Olivares}, {Miret-Roig},
  {Sarro}, {Barrado}, {Berihuete}  \& {Brandner}}{{Galli}
  et~al.}{2020}]{2020A&A...634A..98G}
{Galli} P.~A.~B.,  {Bouy} H.,  {Olivares} J.,  {Miret-Roig} N.,  {Sarro} L.~M.,
   {Barrado} D.,  {Berihuete} A.,   {Brandner} W.,  2020, \mn@doi [\aap]
  {10.1051/0004-6361/201936708}, \href
  {https://ui.adsabs.harvard.edu/abs/2020A&A...634A..98G} {634, A98}

\bibitem[\protect\citeauthoryear{{Galli} et~al.,}{{Galli}
  et~al.}{2021}]{2021A&A...646A..46G}
{Galli} P.~A.~B.,  et~al., 2021, \mn@doi [\aap] {10.1051/0004-6361/202039395},
  \href {https://ui.adsabs.harvard.edu/abs/2021A&A...646A..46G} {646, A46}

\bibitem[\protect\citeauthoryear{{Garcia}, {Hernandez}, {Malaroda}, {Morrell}
  \& {Levato}}{{Garcia} et~al.}{1988}]{1988Ap&SS.148..163G}
{Garcia} B.,  {Hernandez} C.,  {Malaroda} S.,  {Morrell} N.,   {Levato} H.,
  1988, \mn@doi [\apss] {10.1007/BF00646471}, \href
  {https://ui.adsabs.harvard.edu/abs/1988Ap&SS.148..163G} {148, 163}

\bibitem[\protect\citeauthoryear{{Goldman}, {R{\"o}ser}, {Schilbach},
  {Mo{\'o}r}  \& {Henning}}{{Goldman} et~al.}{2018}]{2018ApJ...868...32G}
{Goldman} B.,  {R{\"o}ser} S.,  {Schilbach} E.,  {Mo{\'o}r} A.~C.,   {Henning}
  T.,  2018, \mn@doi [\apj] {10.3847/1538-4357/aae64c}, \href
  {https://ui.adsabs.harvard.edu/abs/2018ApJ...868...32G} {868, 32}

\bibitem[\protect\citeauthoryear{{Gontcharov}}{{Gontcharov}}{2006}]{2006AstL...32..759G}
{Gontcharov} G.~A.,  2006, \mn@doi [Astronomy Letters]
  {10.1134/S1063773706110065}, \href
  {https://ui.adsabs.harvard.edu/abs/2006AstL...32..759G} {32, 759}

\bibitem[\protect\citeauthoryear{{Gonz{\'a}lez} et~al.,}{{Gonz{\'a}lez}
  et~al.}{2021}]{2021A&A...647A..14G}
{Gonz{\'a}lez} M.,  et~al., 2021, \mn@doi [\aap] {10.1051/0004-6361/202038123},
  \href {https://ui.adsabs.harvard.edu/abs/2021A&A...647A..14G} {647, A14}

\bibitem[\protect\citeauthoryear{Green}{Green}{2018}]{Green2018}
Green G.~M.,  2018, \mn@doi [Journal of Open Source Software]
  {10.21105/joss.00695}, 3, 695

\bibitem[\protect\citeauthoryear{{Guenther}, {Esposito}, {Mundt}, {Covino},
  {Alcal{\'a}}, {Cusano}  \& {Stecklum}}{{Guenther}
  et~al.}{2007}]{2007A&A...467.1147G}
{Guenther} E.~W.,  {Esposito} M.,  {Mundt} R.,  {Covino} E.,  {Alcal{\'a}}
  J.~M.,  {Cusano} F.,   {Stecklum} B.,  2007, \mn@doi [\aap]
  {10.1051/0004-6361:20065686}, \href
  {https://ui.adsabs.harvard.edu/abs/2007A&A...467.1147G} {467, 1147}

\bibitem[\protect\citeauthoryear{{Haffert}, {Bohn}, {de Boer}, {Snellen},
  {Brinchmann}, {Girard}, {Keller}  \& {Bacon}}{{Haffert}
  et~al.}{2019}]{2019NatAs...3..749H}
{Haffert} S.~Y.,  {Bohn} A.~J.,  {de Boer} J.,  {Snellen} I.~A.~G.,
  {Brinchmann} J.,  {Girard} J.~H.,  {Keller} C.~U.,   {Bacon} R.,  2019,
  \mn@doi [Nature Astronomy] {10.1038/s41550-019-0780-5}, \href
  {https://ui.adsabs.harvard.edu/abs/2019NatAs...3..749H} {3, 749}

\bibitem[\protect\citeauthoryear{{Hara}, {Tachihara}, {Mizuno}, {Onishi},
  {Kawamura}, {Obayashi}  \& {Fukui}}{{Hara}
  et~al.}{1999}]{1999PASJ...51..895H}
{Hara} A.,  {Tachihara} K.,  {Mizuno} A.,  {Onishi} T.,  {Kawamura} A.,
  {Obayashi} A.,   {Fukui} Y.,  1999, \mn@doi [\pasj] {10.1093/pasj/51.6.895},
  \href {https://ui.adsabs.harvard.edu/abs/1999PASJ...51..895H} {51, 895}

\bibitem[\protect\citeauthoryear{Harris et~al.,}{Harris
  et~al.}{2020}]{harris2020array}
Harris C.~R.,  et~al., 2020, \mn@doi [Nature] {10.1038/s41586-020-2649-2}, 585,
  357

\bibitem[\protect\citeauthoryear{Hunter}{Hunter}{2007}]{doi:10.1109/MCSE.2007.55}
Hunter J.~D.,  2007, \mn@doi [Computing in Science \& Engineering]
  {10.1109/MCSE.2007.55}, 9, 90

\bibitem[\protect\citeauthoryear{{Keppler} et~al.,}{{Keppler}
  et~al.}{2018}]{2018A&A...617A..44K}
{Keppler} M.,  et~al., 2018, \mn@doi [\aap] {10.1051/0004-6361/201832957},
  \href {https://ui.adsabs.harvard.edu/abs/2018A&A...617A..44K} {617, A44}

\bibitem[\protect\citeauthoryear{{Kerr}, {Rizzuto}, {Kraus}  \&
  {Offner}}{{Kerr} et~al.}{2021}]{2021ApJ...917...23K}
{Kerr} R. M.~P.,  {Rizzuto} A.~C.,  {Kraus} A.~L.,   {Offner} S. S.~R.,  2021,
  \mn@doi [\apj] {10.3847/1538-4357/ac0251}, \href
  {https://ui.adsabs.harvard.edu/abs/2021ApJ...917...23K} {917, 23}

\bibitem[\protect\citeauthoryear{{Kharchenko}, {Scholz}, {Piskunov},
  {R{\"o}ser}  \& {Schilbach}}{{Kharchenko} et~al.}{2007}]{2007AN....328..889K}
{Kharchenko} N.~V.,  {Scholz} R.~D.,  {Piskunov} A.~E.,  {R{\"o}ser} S.,
  {Schilbach} E.,  2007, \mn@doi [Astronomische Nachrichten]
  {10.1002/asna.200710776}, \href
  {https://ui.adsabs.harvard.edu/abs/2007AN....328..889K} {328, 889}

\bibitem[\protect\citeauthoryear{{Kounkel} \& {Covey}}{{Kounkel} \&
  {Covey}}{2019}]{2019AJ....158..122K}
{Kounkel} M.,  {Covey} K.,  2019, \mn@doi [\aj] {10.3847/1538-3881/ab339a},
  \href {https://ui.adsabs.harvard.edu/abs/2019AJ....158..122K} {158, 122}

\bibitem[\protect\citeauthoryear{{Kraus} \& {Hillenbrand}}{{Kraus} \&
  {Hillenbrand}}{2008}]{2008ApJ...686L.111K}
{Kraus} A.~L.,  {Hillenbrand} L.~A.,  2008, \mn@doi [\apjl] {10.1086/593012},
  \href {https://ui.adsabs.harvard.edu/abs/2008ApJ...686L.111K} {686, L111}

\bibitem[\protect\citeauthoryear{{Krumholz}, {McKee}  \& {Bland
  -Hawthorn}}{{Krumholz} et~al.}{2019}]{Krumholz19a}
{Krumholz} M.~R.,  {McKee} C.~F.,   {Bland -Hawthorn} J.,  2019, \mn@doi
  [\araa] {10.1146/annurev-astro-091918-104430}, \href
  {https://ui.adsabs.harvard.edu/abs/2019ARA&A..57..227K} {57, 227}

\bibitem[\protect\citeauthoryear{{Leike}, {Glatzle}  \& {En{\ss}lin}}{{Leike}
  et~al.}{2020}]{2020A&A...639A.138L}
{Leike} R.~H.,  {Glatzle} M.,   {En{\ss}lin} T.~A.,  2020, \mn@doi [\aap]
  {10.1051/0004-6361/202038169}, \href
  {https://ui.adsabs.harvard.edu/abs/2020A&A...639A.138L} {639, A138}

\bibitem[\protect\citeauthoryear{{Lopez Mart{\'\i}}, {Jimenez Esteban}, {Bayo},
  {Barrado}, {Solano}  \& {Rodrigo}}{{Lopez Mart{\'\i}}
  et~al.}{2013}]{2013A&A...551A..46L}
{Lopez Mart{\'\i}} B.,  {Jimenez Esteban} F.,  {Bayo} A.,  {Barrado} D.,
  {Solano} E.,   {Rodrigo} C.,  2013, \mn@doi [\aap]
  {10.1051/0004-6361/201220128}, \href
  {https://ui.adsabs.harvard.edu/abs/2013A&A...551A..46L} {551, A46}

\bibitem[\protect\citeauthoryear{{Luhman} \& {Esplin}}{{Luhman} \&
  {Esplin}}{2020}]{2020AJ....160...44L}
{Luhman} K.~L.,  {Esplin} T.~L.,  2020, \mn@doi [\aj]
  {10.3847/1538-3881/ab9599}, \href
  {https://ui.adsabs.harvard.edu/abs/2020AJ....160...44L} {160, 44}

\bibitem[\protect\citeauthoryear{{Makarov}, {Olling}  \& {Teuben}}{{Makarov}
  et~al.}{2004}]{2004MNRAS.352.1199M}
{Makarov} V.~V.,  {Olling} R.~P.,   {Teuben} P.~J.,  2004, \mn@doi [\mnras]
  {10.1111/j.1365-2966.2004.08012.x}, \href
  {https://ui.adsabs.harvard.edu/abs/2004MNRAS.352.1199M} {352, 1199}

\bibitem[\protect\citeauthoryear{{Mamajek} \& {Bell}}{{Mamajek} \&
  {Bell}}{2014}]{2014MNRAS.445.2169M}
{Mamajek} E.~E.,  {Bell} C. P.~M.,  2014, \mn@doi [\mnras]
  {10.1093/mnras/stu1894}, \href
  {https://ui.adsabs.harvard.edu/abs/2014MNRAS.445.2169M} {445, 2169}

\bibitem[\protect\citeauthoryear{{Mamajek}, {Meyer}  \& {Liebert}}{{Mamajek}
  et~al.}{2002}]{2002AJ....124.1670M}
{Mamajek} E.~E.,  {Meyer} M.~R.,   {Liebert} J.,  2002, \mn@doi [\aj]
  {10.1086/341952}, \href
  {https://ui.adsabs.harvard.edu/abs/2002AJ....124.1670M} {124, 1670}

\bibitem[\protect\citeauthoryear{{Mann} et~al.,}{{Mann}
  et~al.}{2021}]{2021arXiv211009531M}
{Mann} A.~W.,  et~al., 2021, arXiv e-prints, \href
  {https://ui.adsabs.harvard.edu/abs/2021arXiv211009531M} {p. arXiv:2110.09531}

\bibitem[\protect\citeauthoryear{{Mermilliod}, {Mayor}  \& {Udry}}{{Mermilliod}
  et~al.}{2009}]{2009A&A...498..949M}
{Mermilliod} J.~C.,  {Mayor} M.,   {Udry} S.,  2009, \mn@doi [\aap]
  {10.1051/0004-6361/200810244}, \href
  {https://ui.adsabs.harvard.edu/abs/2009A&A...498..949M} {498, 949}

\bibitem[\protect\citeauthoryear{{Miret-Roig} et~al.,}{{Miret-Roig}
  et~al.}{2020}]{2020A&A...642A.179M}
{Miret-Roig} N.,  et~al., 2020, \mn@doi [\aap] {10.1051/0004-6361/202038765},
  \href {https://ui.adsabs.harvard.edu/abs/2020A&A...642A.179M} {642, A179}

\bibitem[\protect\citeauthoryear{{M{\"u}ller} et~al.,}{{M{\"u}ller}
  et~al.}{2018}]{2018A&A...617L...2M}
{M{\"u}ller} A.,  et~al., 2018, \mn@doi [\aap] {10.1051/0004-6361/201833584},
  \href {https://ui.adsabs.harvard.edu/abs/2018A&A...617L...2M} {617, L2}

\bibitem[\protect\citeauthoryear{{Murphy}, {Lawson}  \& {Bessell}}{{Murphy}
  et~al.}{2013}]{2013MNRAS.435.1325M}
{Murphy} S.~J.,  {Lawson} W.~A.,   {Bessell} M.~S.,  2013, \mn@doi [\mnras]
  {10.1093/mnras/stt1375}, \href
  {https://ui.adsabs.harvard.edu/abs/2013MNRAS.435.1325M} {435, 1325}

\bibitem[\protect\citeauthoryear{{Nelder} \& {Mead}}{{Nelder} \&
  {Mead}}{1965}]{NelderMead}
{Nelder} J.~A.,  {Mead} R.,  1965, \mn@doi [The Computer Journal]
  {10.1093/comjnl/7.4.308}, 7, 308

\bibitem[\protect\citeauthoryear{{Neuh{\"a}user} et~al.,}{{Neuh{\"a}user}
  et~al.}{2000}]{2000A&AS..146..323N}
{Neuh{\"a}user} R.,  et~al., 2000, \mn@doi [\aaps] {10.1051/aas:2000272}, \href
  {https://ui.adsabs.harvard.edu/abs/2000A&AS..146..323N} {146, 323}

\bibitem[\protect\citeauthoryear{{Pecaut} \& {Mamajek}}{{Pecaut} \&
  {Mamajek}}{2013}]{2013ApJS..208....9P}
{Pecaut} M.~J.,  {Mamajek} E.~E.,  2013, \mn@doi [\apjs]
  {10.1088/0067-0049/208/1/9}, \href
  {http://adsabs.harvard.edu/abs/2013ApJS..208....9P} {208, 9}

\bibitem[\protect\citeauthoryear{{Pecaut} \& {Mamajek}}{{Pecaut} \&
  {Mamajek}}{2016}]{2016MNRAS.461..794P}
{Pecaut} M.~J.,  {Mamajek} E.~E.,  2016, \mn@doi [\mnras]
  {10.1093/mnras/stw1300}, \href
  {https://ui.adsabs.harvard.edu/abs/2016MNRAS.461..794P} {461, 794}

\bibitem[\protect\citeauthoryear{{Pecaut}, {Mamajek}  \& {Bubar}}{{Pecaut}
  et~al.}{2012}]{2012ApJ...746..154P}
{Pecaut} M.~J.,  {Mamajek} E.~E.,   {Bubar} E.~J.,  2012, \mn@doi [\apj]
  {10.1088/0004-637X/746/2/154}, \href
  {https://ui.adsabs.harvard.edu/abs/2012ApJ...746..154P} {746, 154}

\bibitem[\protect\citeauthoryear{P\'erez \& Granger}{P\'erez \&
  Granger}{2007}]{doi:10.1109/MCSE.2007.53}
P\'erez F.,  Granger B.~E.,  2007, \mn@doi [Computing in Science \&
  Engineering] {10.1109/MCSE.2007.53}, 9, 21

\bibitem[\protect\citeauthoryear{{Platais}, {Kozhurina-Platais}  \& {van
  Leeuwen}}{{Platais} et~al.}{1998}]{1998AJ....116.2423P}
{Platais} I.,  {Kozhurina-Platais} V.,   {van Leeuwen} F.,  1998, \mn@doi [\aj]
  {10.1086/300606}, \href
  {https://ui.adsabs.harvard.edu/abs/1998AJ....116.2423P} {116, 2423}

\bibitem[\protect\citeauthoryear{{Prato}}{{Prato}}{2007}]{2007ApJ...657..338P}
{Prato} L.,  2007, \mn@doi [\apj] {10.1086/510882}, \href
  {https://ui.adsabs.harvard.edu/abs/2007ApJ...657..338P} {657, 338}

\bibitem[\protect\citeauthoryear{{Preibisch} \& {Mamajek}}{{Preibisch} \&
  {Mamajek}}{2008}]{2008hsf2.book..235P}
{Preibisch} T.,  {Mamajek} E.,  2008, {The Nearest OB Association:
  Scorpius-Centaurus (Sco OB2)}.
p.~235

\bibitem[\protect\citeauthoryear{{Preibisch}, {Brown}, {Bridges}, {Guenther}
  \& {Zinnecker}}{{Preibisch} et~al.}{2002}]{2002AJ....124..404P}
{Preibisch} T.,  {Brown} A. G.~A.,  {Bridges} T.,  {Guenther} E.,   {Zinnecker}
  H.,  2002, \mn@doi [\aj] {10.1086/341174}, \href
  {https://ui.adsabs.harvard.edu/abs/2002AJ....124..404P} {124, 404}

\bibitem[\protect\citeauthoryear{{Price-Whelan} et~al.,}{{Price-Whelan}
  et~al.}{2018}]{astropy:2018}
{Price-Whelan} A.~M.,  et~al., 2018, \mn@doi [\aj] {10.3847/1538-3881/aabc4f},
  \href {https://ui.adsabs.harvard.edu/#abs/2018AJ....156..123T} {156, 123}

\bibitem[\protect\citeauthoryear{{Rains} et~al.,}{{Rains}
  et~al.}{2021}]{2021MNRAS.504.5788R}
{Rains} A.~D.,  et~al., 2021, \mn@doi [\mnras] {10.1093/mnras/stab1167}, \href
  {https://ui.adsabs.harvard.edu/abs/2021MNRAS.504.5788R} {504, 5788}

\bibitem[\protect\citeauthoryear{{Riedel}, {Blunt}, {Lambrides}, {Rice}, {Cruz}
   \& {Faherty}}{{Riedel} et~al.}{2017}]{2017AJ....153...95R}
{Riedel} A.~R.,  {Blunt} S.~C.,  {Lambrides} E.~L.,  {Rice} E.~L.,  {Cruz}
  K.~L.,   {Faherty} J.~K.,  2017, \mn@doi [\aj] {10.3847/1538-3881/153/3/95},
  \href {https://ui.adsabs.harvard.edu/abs/2017AJ....153...95R} {153, 95}

\bibitem[\protect\citeauthoryear{{Rizzuto}, {Ireland}  \&
  {Robertson}}{{Rizzuto} et~al.}{2011}]{2011MNRAS.416.3108R}
{Rizzuto} A.~C.,  {Ireland} M.~J.,   {Robertson} J.~G.,  2011, \mn@doi [\mnras]
  {10.1111/j.1365-2966.2011.19256.x}, \href
  {https://ui.adsabs.harvard.edu/abs/2011MNRAS.416.3108R} {416, 3108}

\bibitem[\protect\citeauthoryear{{Rizzuto}, {Ireland}  \& {Kraus}}{{Rizzuto}
  et~al.}{2015}]{2015MNRAS.448.2737R}
{Rizzuto} A.~C.,  {Ireland} M.~J.,   {Kraus} A.~L.,  2015, \mn@doi [\mnras]
  {10.1093/mnras/stv207}, \href
  {https://ui.adsabs.harvard.edu/abs/2015MNRAS.448.2737R} {448, 2737}

\bibitem[\protect\citeauthoryear{{Rizzuto}, {Ireland}, {Dupuy}  \&
  {Kraus}}{{Rizzuto} et~al.}{2016}]{2016ApJ...817..164R}
{Rizzuto} A.~C.,  {Ireland} M.~J.,  {Dupuy} T.~J.,   {Kraus} A.~L.,  2016,
  \mn@doi [\apj] {10.3847/0004-637X/817/2/164}, \href
  {https://ui.adsabs.harvard.edu/abs/2016ApJ...817..164R} {817, 164}

\bibitem[\protect\citeauthoryear{{R{\"o}ser}, {Schilbach}, {Goldman},
  {Henning}, {Moor}  \& {Derekas}}{{R{\"o}ser}
  et~al.}{2018}]{2018A&A...614A..81R}
{R{\"o}ser} S.,  {Schilbach} E.,  {Goldman} B.,  {Henning} T.,  {Moor} A.,
  {Derekas} A.,  2018, \mn@doi [\aap] {10.1051/0004-6361/201732213}, \href
  {https://ui.adsabs.harvard.edu/abs/2018A&A...614A..81R} {614, A81}

\bibitem[\protect\citeauthoryear{{Sartoretti} et~al.,}{{Sartoretti}
  et~al.}{2018}]{2018A&A...616A...6S}
{Sartoretti} P.,  et~al., 2018, \mn@doi [\aap] {10.1051/0004-6361/201832836},
  \href {https://ui.adsabs.harvard.edu/abs/2018A&A...616A...6S} {616, A6}

\bibitem[\protect\citeauthoryear{{Sch{\"o}nrich}, {Binney}  \&
  {Dehnen}}{{Sch{\"o}nrich} et~al.}{2010}]{2010MNRAS.403.1829S}
{Sch{\"o}nrich} R.,  {Binney} J.,   {Dehnen} W.,  2010, \mn@doi [\mnras]
  {10.1111/j.1365-2966.2010.16253.x}, \href
  {https://ui.adsabs.harvard.edu/abs/2010MNRAS.403.1829S} {403, 1829}

\bibitem[\protect\citeauthoryear{{Schutz}, {Lin}, {Safdi}  \& {Wu}}{{Schutz}
  et~al.}{2018}]{2018PhRvL.121h1101S}
{Schutz} K.,  {Lin} T.,  {Safdi} B.~R.,   {Wu} C.-L.,  2018, \mn@doi [\prl]
  {10.1103/PhysRevLett.121.081101}, \href
  {https://ui.adsabs.harvard.edu/abs/2018PhRvL.121h1101S} {121, 081101}

\bibitem[\protect\citeauthoryear{{Siebert} et~al.,}{{Siebert}
  et~al.}{2011}]{2011AJ....141..187S}
{Siebert} A.,  et~al., 2011, \mn@doi [\aj] {10.1088/0004-6256/141/6/187}, \href
  {https://ui.adsabs.harvard.edu/abs/2011AJ....141..187S} {141, 187}

\bibitem[\protect\citeauthoryear{{Squicciarini}, {Gratton}, {Bonavita}  \&
  {Mesa}}{{Squicciarini} et~al.}{2021}]{2021MNRAS.tmp.1862S}
{Squicciarini} V.,  {Gratton} R.,  {Bonavita} M.,   {Mesa} D.,  2021, \mn@doi
  [\mnras] {10.1093/mnras/stab2079}, \href
  {https://ui.adsabs.harvard.edu/abs/2021MNRAS.tmp.1862S} {}

\bibitem[\protect\citeauthoryear{{Steinmetz} et~al.,}{{Steinmetz}
  et~al.}{2020}]{2020AJ....160...82S}
{Steinmetz} M.,  et~al., 2020, \mn@doi [\aj] {10.3847/1538-3881/ab9ab9}, \href
  {https://ui.adsabs.harvard.edu/abs/2020AJ....160...82S} {160, 82}

\bibitem[\protect\citeauthoryear{{Sullivan} \& {Kraus}}{{Sullivan} \&
  {Kraus}}{2021}]{2021ApJ...912..137S}
{Sullivan} K.,  {Kraus} A.~L.,  2021, \mn@doi [\apj]
  {10.3847/1538-4357/abf044}, \href
  {https://ui.adsabs.harvard.edu/abs/2021ApJ...912..137S} {912, 137}

\bibitem[\protect\citeauthoryear{{Thanathibodee} et~al.,}{{Thanathibodee}
  et~al.}{2020}]{2020ApJ...892...81T}
{Thanathibodee} T.,  et~al., 2020, \mn@doi [\apj] {10.3847/1538-4357/ab77c1},
  \href {https://ui.adsabs.harvard.edu/abs/2020ApJ...892...81T} {892, 81}

\bibitem[\protect\citeauthoryear{{Torres}, {Quast}, {da Silva}, {de La Reza},
  {Melo}  \& {Sterzik}}{{Torres} et~al.}{2006}]{2006A&A...460..695T}
{Torres} C.~A.~O.,  {Quast} G.~R.,  {da Silva} L.,  {de La Reza} R.,  {Melo}
  C.~H.~F.,   {Sterzik} M.,  2006, \mn@doi [\aap] {10.1051/0004-6361:20065602},
  \href {https://ui.adsabs.harvard.edu/abs/2006A&A...460..695T} {460, 695}

\bibitem[\protect\citeauthoryear{{Wheeler}, {Hogg}  \& {Ness}}{{Wheeler}
  et~al.}{2021}]{2021ApJ...908..247W}
{Wheeler} A.~J.,  {Hogg} D.~W.,   {Ness} M.,  2021, \mn@doi [\apj]
  {10.3847/1538-4357/abd544}, \href
  {https://ui.adsabs.harvard.edu/abs/2021ApJ...908..247W} {908, 247}

\bibitem[\protect\citeauthoryear{{White}, {Gabor}  \& {Hillenbrand}}{{White}
  et~al.}{2007}]{2007AJ....133.2524W}
{White} R.~J.,  {Gabor} J.~M.,   {Hillenbrand} L.~A.,  2007, \mn@doi [\aj]
  {10.1086/514336}, \href
  {https://ui.adsabs.harvard.edu/abs/2007AJ....133.2524W} {133, 2524}

\bibitem[\protect\citeauthoryear{{Wilson}}{{Wilson}}{1953}]{1953GCRV..C......0W}
{Wilson} R.~E.,  1953, Carnegie Institute Washington D.C. Publication, \href
  {https://ui.adsabs.harvard.edu/abs/1953GCRV..C......0W} {p.~0}

\bibitem[\protect\citeauthoryear{{Wright} \& {Mamajek}}{{Wright} \&
  {Mamajek}}{2018}]{2018MNRAS.476..381W}
{Wright} N.~J.,  {Mamajek} E.~E.,  2018, \mn@doi [\mnras]
  {10.1093/mnras/sty207}, \href
  {https://ui.adsabs.harvard.edu/abs/2018MNRAS.476..381W} {476, 381}

\bibitem[\protect\citeauthoryear{{Yen}, {Reffert}, {Schilbach}, {R{\"o}ser},
  {Kharchenko}  \& {Piskunov}}{{Yen} et~al.}{2018}]{2018A&A...615A..12Y}
{Yen} S.~X.,  {Reffert} S.,  {Schilbach} E.,  {R{\"o}ser} S.,  {Kharchenko}
  N.~V.,   {Piskunov} A.~E.,  2018, \mn@doi [\aap]
  {10.1051/0004-6361/201731905}, \href
  {https://ui.adsabs.harvard.edu/abs/2018A&A...615A..12Y} {615, A12}

\bibitem[\protect\citeauthoryear{{de Geus}, {de Zeeuw}  \& {Lub}}{{de Geus}
  et~al.}{1989}]{1989A&A...216...44D}
{de Geus} E.~J.,  {de Zeeuw} P.~T.,   {Lub} J.,  1989, \aap, \href
  {https://ui.adsabs.harvard.edu/abs/1989A&A...216...44D} {216, 44}

\bibitem[\protect\citeauthoryear{{de Zeeuw}, {Hoogerwerf}, {de Bruijne},
  {Brown}  \& {Blaauw}}{{de Zeeuw} et~al.}{1999}]{1999AJ....117..354D}
{de Zeeuw} P.~T.,  {Hoogerwerf} R.,  {de Bruijne} J.~H.~J.,  {Brown} A.~G.~A.,
   {Blaauw} A.,  1999, \mn@doi [\aj] {10.1086/300682}, \href
  {https://ui.adsabs.harvard.edu/abs/1999AJ....117..354D} {117, 354}

\bibitem[\protect\citeauthoryear{{{\v{Z}}erjal} et~al.,}{{{\v{Z}}erjal}
  et~al.}{2019}]{2019MNRAS.484.4591Z}
{{\v{Z}}erjal} M.,  et~al., 2019, \mn@doi [\mnras] {10.1093/mnras/stz296},
  \href {https://ui.adsabs.harvard.edu/abs/2019MNRAS.484.4591Z} {484, 4591}

\bibitem[\protect\citeauthoryear{{{\v{Z}}erjal} et~al.,}{{{\v{Z}}erjal}
  et~al.}{2021}]{2021MNRAS.503..938Z}
{{\v{Z}}erjal} M.,  et~al., 2021, \mn@doi [\mnras] {10.1093/mnras/stab513},
  \href {https://ui.adsabs.harvard.edu/abs/2021MNRAS.503..938Z} {503, 938}

\makeatother
\end{thebibliography}
